\documentclass[%
12pt,
aip,
cha,
amsmath,amssymb,
reprint,
]{revtex4-1}

\usepackage{graphicx}
\usepackage{dcolumn}
\usepackage{bm}
\usepackage[utf8]{inputenc}
\usepackage[T1]{fontenc}
\usepackage{xcolor}

\def\pct{\%}
\renewcommand{\d}[1]{\ensuremath{\operatorname{d}\!{#1}}}
\renewcommand{\Pr}{\mathop{\rm prob}\nolimits}
\DeclareMathOperator*{\m}{area}
\DeclareMathOperator*{\argmax}{arg\,max}

\definecolor{xkcd:purple}{HTML}{7e1e9c}
\definecolor{xkcd:green}{HTML}{15b01a}
\definecolor{xkcd:blue}{HTML}{0343df}
\definecolor{xkcd:pink}{HTML}{ff81c0}
\definecolor{xkcd:brown}{HTML}{653700}
\definecolor{xkcd:red}{HTML}{e50000}
\definecolor{xkcd:lightblue}{HTML}{95d0fc}
\definecolor{xkcd:teal}{HTML}{029386}
\definecolor{xkcd:orange}{HTML}{f97306}
\definecolor{xkcd:lightgreen}{HTML}{96f97b}
\definecolor{xkcd:majenta}{HTML}{c20078}
\definecolor{xkcd:grey}{HTML}{929591}
\definecolor{xkcd:lightpurple}{HTML}{bf77f6}
\definecolor{xkcd:turquoise}{HTML}{06c2ac}
\definecolor{xkcd:tan}{HTML}{d1b26f}
\definecolor{xkcd:olive}{HTML}{6e750e}
\definecolor{xkcd:salmon}{HTML}{ff796c}

\begin{document}

\title{Markov-chain-inspired search for MH370}

\author{P.\ Miron}
\email{pmiron@miami.edu}
\affiliation{Department of Atmospheric Sciences, Rosenstiel School of
Marine and Atmospheric Science, University of Miami, Miami, Florida,
USA} 

\author{F.\ J.\ Beron-Vera}
\affiliation{Department of Atmospheric Sciences, Rosenstiel School of
Marine and Atmospheric Science, University of Miami, Miami, Florida,
USA} 

\author{M.\ J.\ Olascoaga}
\affiliation{Department of Ocean Sciences, Rosenstiel School of Marine
and Atmospheric Science, University of Miami, Miami, Florida, USA}

\author{P.\ Koltai}
\affiliation{Institute of Mathematics, Freie Universit\"at Berlin, Berlin, Germany}

\date{\today}

\begin{abstract}
Markov-chain models are constructed for the probabilistic description
of the drift of marine debris from Malaysian Airlines flight MH370.
En route from Kuala Lumpur to Beijing, the MH370 mysteriously
disappeared in the southeastern Indian Ocean on 8 March 2014,
somewhere along the arc of the 7th ping ring around the Inmarsat-3F1
satellite position when the airplane lost contact.  The models are
obtained by discretizing the motion of undrogued satellite-tracked
surface drifting buoys from the global historical data bank. A
spectral analysis, Bayesian estimation, and the computation of most
probable paths between the Inmarsat arc and confirmed airplane
debris beaching sites are shown to constrain the crash site, near
25$^{\circ}$S on the Inmarsat arc.
\end{abstract}

\pacs{02.50.Ga; 47.27.De; 92.10.Fj}

\maketitle

\begin{quotation}
Application of tools from ergodic theory on historical Lagrangian
ocean data is shown to constrain the crash site of Malaysian Airlines
flight MH370 given airplane debris beaching site information.  The
disappearance of flight MH370 constitutes one of the most enigmatic
episodes in the history of commercial aviation.  The tools employed
have far-reaching applicability as they are particularly well suited
in inverse modeling, critical for instance in revealing contamination
sources in the ocean and the atmosphere.  The only requirement for
their success is sufficient spatiotemporal Lagrangian sampling.
\end{quotation}

\section{Introduction}

The disappearance in the southeastern Indian ocean on 8 March 2014
of Malaysian Airlines flight MH370 en route from Kuala Lumpur to
Beijing is one of the biggest aviation mysteries. With the loss of
all 227 passengers and 12 crew members on board, flight MH370 is
the second deadliest incident involving a Boeing 777 aircraft.  At
a cost nearing \$155 million its search is already the most expensive
in aviation history.

In January 2017, almost three years after the airplane disappearance,
the Australian Government's Joint Agency Coordination Centre halted
the search after failing to locate the airplane across more than
120,000 km$^2$ in the eastern Indian Ocean.  On May 29th 2018, the
latest attempt to locate the aircraft completed after an unsuccessful
several-month cruise by ocean exploration company Ocean Infinity
through an agreement with the Malaysian Government.

Analysis \cite{Ashton-etal-15, Holland-18} of the Inmarsat-3F1
satellite communication, provided in the form of handshakes between
engines and satellite, indicated that the aircraft had lost contact
along the 7th ping ring around the position of the satellite on 8
March 2014, ranging from Java, Indonesia, to the southern Indian
Ocean, southwest of Australia (Fig.\ \ref{fig:grid}).  Since then,
several pieces of marine debris belonging to the airplane have been
found washed up on the shores of various coastlines in the southwestern
Indian Ocean (Fig.\ \ref{fig:grid}).  The first debris piece was
discovered on 29 July 2015 on a beach of Reunion Island.  Locations
and dates from eight confirmed beachings \cite{ATBS-18} are indicated
in Table \ref{tab:debris}.

\begin{table*}
  \centering
  \begin{tabular}{lcc}
  \hline%
  \textbf{Beaching site} & \textbf{Days since crash} &
  \textbf{Color in plots}\\
  Reunion Island (RE) & 508 & 
  \raisebox{3pt}{\fcolorbox{white}{xkcd:purple}{\rule{0pt}{.5pt}\rule{.5pt}{0pt}}}\\
  South Africa (ZA) & 655 & 
  \raisebox{3pt}{\fcolorbox{white}{xkcd:green}{\rule{0pt}{.5pt}\rule{.5pt}{0pt}}}\\
  Mozambique (MZ) & 662 & 
  \raisebox{3pt}{\fcolorbox{white}{xkcd:blue}{\rule{0pt}{.5pt}\rule{.5pt}{0pt}}}\\
  Mozambique (MZ) & 721 & 
  \raisebox{3pt}{\fcolorbox{white}{xkcd:pink}{\rule{0pt}{.5pt}\rule{.5pt}{0pt}}}\\
  Mauritius (MU) & 753 & 
  \raisebox{3pt}{\fcolorbox{white}{xkcd:brown}{\rule{0pt}{.5pt}\rule{.5pt}{0pt}}}\\
  Madagascar (MG) & 808 & 
  \raisebox{3pt}{\fcolorbox{white}{xkcd:red}{\rule{0pt}{.5pt}\rule{0.5pt}{0pt}}}\\
  Mauritius/Rodrigues (MU/RRG) & 827 & 
  \raisebox{3pt}{\fcolorbox{white}{xkcd:lightblue}{\rule{0pt}{.5pt}\rule{.5pt}{0pt}}}\\
  Tanzania (TZ) & 838 & 
  \raisebox{3pt}{\fcolorbox{white}{xkcd:teal}{\rule{0pt}{.5pt}\rule{.5pt}{0pt}}}\\
  \hline%
  \end{tabular}
  \caption{Beaching information for confirmed debris from Malaysian
  Airlines flight MH370, which crashed in the Indian Ocean on 8
  March 2014.} 
  \label{tab:debris}
\end{table*}

MH370 search approaches to date have included: the application of
geometric nonlinear dynamics tools on simulated surface flows or
as inferred from satellite altimetry observations with a focus on
the analysis of the efficacy of the aerial search \cite{Garcia-etal-15};
attempts to backward trajectory reconstruction from drifter relative
dispersion properties \cite{Corrado-etal-17}; inspection of
trajectories of satellite-tracked surface drifting buoys (drifters)
and direct trajectory forward integrations of altimetry-derived
currents corrected using drifter velocities \cite{Trinanes-etal-16};
forward trajectory integrations of various flow representations
corrected to account for leeway drift \cite{Nesterov-18,
Griffin-etal-17, Maximenko-etal-15, vanOrmondt-Baart-15}; Bayesian
inference of debris beaching sites using trajectories from an
ensemble of model velocity realizations \cite{Jansen-etal-16};
backward trajectory integrations of simulated velocities
\cite{Durgadoo-Biastoch-15}; consideration of Bayesian methods for
estimating commercial aircraft trajectories using models of the
information contained in satellite communications messages and of
the aircraft dynamics \cite{Davey-etal-16}; and biochemical analysis
of barnacles attached to debris washed ashore to infer the temperature
of the water they were exposed to \cite{deDeckker-17}.

Here we introduce a novel framework for locating the MH370 crash
site.  Rooted in probabilistic nonlinear dynamical systems theory,
the framework uses the locations and times of confirmed airplane
debris beachings and historical trajectories produced by drifters
to restrict the crash site along the Inmarsat arc.  An additional
important aspect of our approach to MH370 search, which makes it
quite different than the previous ones, is that it directly targets
crash site localization while exclusively performing forward
evolutions.

Even though we have chosen to carry out a fully data-based analysis,
the framework may be applied on the output from a data-assimilative
system, enabling operational use of it in guiding search efforts,
currently suspended, if they ever are to be resumed

\section{Setup}

The probabilistic framework to be developed in this paper builds
on well-established results from ergodic theory \cite{Lasota-Mackey-94}
and Markov chains \cite{Bremaud-99, Norris-98}, which place the
focus on the evolution of probability densities rather than individual
trajectories in the phase space of a nonlinear dynamical system
(mathematical details are deferred to Appendix A in the Supplementary
Material).  At the core of the measure-theoretic characterization
of nonlinear dynamics is the transfer operator and its discrete
version, the transition matrix.  The relevant dynamical system here
is that obeyed by trajectories of airplane debris pieces that are
transported under the combined action of turbulent ocean currents
and winds mediated by inertia \cite{Beron-etal-15, Beron-etal-16}.

Let $\{\xi_{t+kT}\}_{k\ge 0}$ denote the time-discrete stochastic
process describing such random trajectories.  Assuming that this
process is time-homogeneous over a sufficient long time interval
$\mathfrak T$, its transition probabilities are described by a
stochastic kernel, say $K(x,y) \ge 0$ such that $\smash{\int_X
K(x,y)\d{y} = 1}$ for all $x$ in phase space $X$, represented by
the surface-ocean domain of interest. A probability density $f(x)
\ge 0 $, $\smash{\int_X f(x) \d{x} = 1}$, describing the distribution
of $\xi_t$ at any time $t\in \mathfrak T$ evolves to
\begin{equation}
   \mathcal Pf(y) = \int_X K(x,y)f(x)\d{x}
\end{equation}
at time $t+T\in \mathfrak T$, which defines a Markov operator
$\mathcal P : L^1(X) \circlearrowleft$ generally known as a \emph{transfer
operator} \cite{Lasota-Mackey-94}.

To infer the action of $\mathcal P$ from a discrete set of trajectories
one can use a Galerkin approximation referred to as Ulam's method
\cite{Ulam-60, Kovacs-Tel-89, Koltai-10}.  This approach consists
in covering $X$ with $N$ connected boxes $\{B_1, \dotsc, B_N\}$,
disjoint up to zero-measure intersections, and projecting functions
in $L^1(X)$ onto a finite-dimensional space $V_N$ spanned by indicator
functions of boxes normalized by box area.  The discrete action of
$\mathcal P$ on $V_N$ is described by a matrix $P \in \mathbb{R}^{N\times
N}$ called a \emph{transition matrix}.  Let $\xi_t$ be a position
chosen at random from a uniform distribution on $B_i$ at time $t$.
Then
\begin{equation}
  P_{ij} = \Pr[\xi_{t+T}\in B_j \mid \xi_t \in B_i] =
  \frac{\int_{B_i} \int_{\!{B_j}} K(x,y) \d{x}\d{y}}{\m(B_i)},
\end{equation}
which can be estimated as (cf.\ Appendix A of Miron et al.\
\cite{Miron-etal-18a})
\begin{equation}
  P_{ij} \approx \frac{\mbox{\# points in $B_i$ at $t$ that
  evolve to $B_j$ at $t+T$}}{\mbox{\# points in $B_i$ at $t$}}. 
  \label{eq:P}
\end{equation}
Note that $\sum_j P_{ij} = 1$ for all $i$, so $P$ is a (row)
stochastic matrix that defines a \emph{Markov chain} on boxes, which
represent the states of the chain \cite{Bremaud-99, Norris-98}.
The evolution of the discrete representation of $f(x)$, i.e., a
probability vector $\mathbf{f} = (f_1, \cdots ,f_N)$, $\smash{\sum
f_i = 1}$ where $f_i = \int_{B_i} f(x)\d{x}$, is calculated under
left multiplication, i.e.,
\begin{equation}
  \mathbf{f}^{(k)} = \mathbf{f}P^{k},\quad k = 1, 2, \dotsc.
  \label{eq:pP}
\end{equation}

\section{Construction of a suitable transition matrix}

The surface circulation of the Indian Ocean is influenced by monsoon
intraseasonal variability \cite{Schott-McCreary-01}.  An appropriate
Markov-chain model for marine debris motion in the Indian Ocean
must account for this variability, which we do in constructing the
chain's $P$ in a fully data-based fashion using trajectories produced
by satellite-tracked drifters.

The drifter data are collected by the National Oceanic Atmospheric
Administration/Global Drifter Program (NOAA/GDP) \cite{Lumpkin-Pazos-07}.
Trajectories sampling the world oceans including the Indian Ocean
exist since 1979.  For the purpose of the present analysis we
restrict attention to trajectory portions during which the drogue
(a 15-m-long sea anchor designed \cite{Niiler-Paduan-95} to minimize
wind slippage and wave-induced drift) attached to the spherical
float carrying the satellite tracker is absent \cite{Lumpkin-etal-12}.
Undrogued drifter motion is affected by inertial effects (i.e.,
those produced by buoyancy, finite-size, and shape) and thus is
more representative of marine debris motion than that of drogued
drifters, which more closely follow water motion \cite{Beron-etal-16}.

\begin{figure}[t!]
  \centering
  \includegraphics[width=\columnwidth]{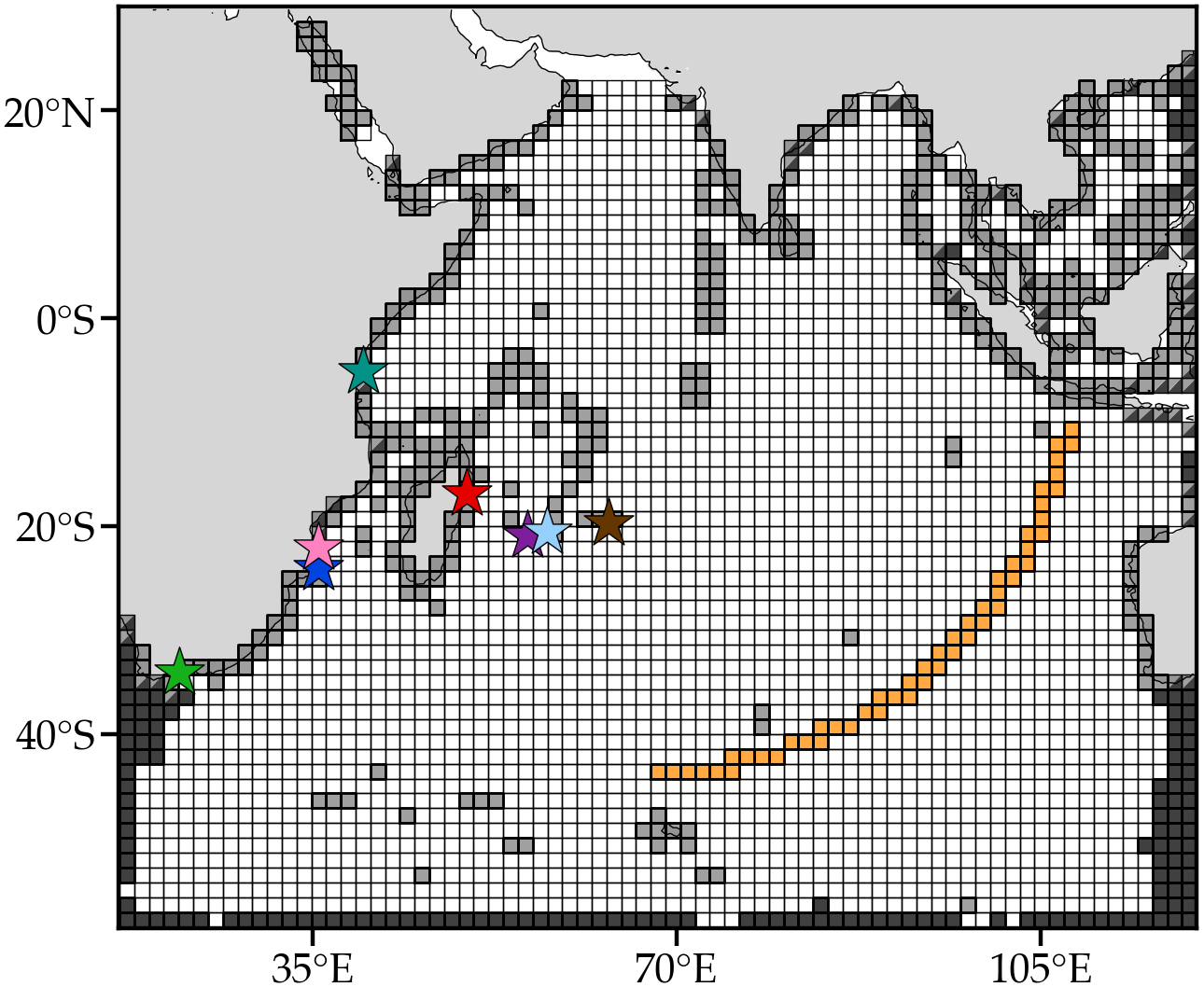}%
  \caption{Covering of the Indian Ocean domain into boxes forming
  the various Markov-chain models constructed using satellite-tracked
  undrogued drifters to describe the motion of marine debris produced
  by the crash of Malaysian Airlines flight MH370. Boxes with
  positive probability of the chain(s) to terminate outside the
  domain (leaky states) are indicated in dark gray, boxes including
  land--water interfaces (sticky states) are shown in light gray,
  and boxes along an arc of 7th ping ring around the Inmarsat-3F1
  satellite position when communication with the MH370 flight was
  lost (crash states) are highlighted in yellow. Stars correspond
  to the airplane debris beaching sites in Table \ref{tab:debris}.}
  \label{fig:grid}%
\end{figure}

To construct $P$, we cover the Indian Ocean domain with a grid of
0.25$^{\circ}$ $\times$ 0.25$^{\circ}$ longitude--latitude boxes
(Fig.\ \ref{fig:grid}).  The size of the cells was selected to
maximize the grid's resolution while each individual box is sampled
by enough trajectories.  Similar grid resolutions in analysis
involving buoy trajectory data were employed in recent work
\cite{Miron-etal-17, Miron-etal-18a, Olascoaga-etal-18}, where
sensitivity analyses to cell size variations and data amount
truncations are presented.  The area of the boxes varies from about
400 to 750 km$^2$, yet the normalization by box area in the definition
of the vector space $V_N$ makes this variation inconsequential,
i.e., a stochastic transition matrix is obtained without the need
of a similarity transformation (e.g., Froyland and Padberg
\cite{Froyland-Padberg-09}).  Ignoring time, there are 226 drifters
on average per box (cf.\ Fig.\ S1 in the Supplementary Material);
the number of drifters vary between 36 and 58 if the data are grouped
according to season of the year. Equation (\ref{eq:P}) is then
evaluated for appropriate transition time ($T$) and time-homogeneity
interval ($\mathfrak T$) choices.

Using $T = 1$ d, approximately the surface ocean Lagrangian
decorrelation time \cite{Lacasce-08}, the simple Markovian dynamics
test $\lambda(P(nT)) = \lambda(P(T))^n$, where $\lambda$ denotes
eigenvalue, holds very well up to $n = 10$ (cf.\ Fig.\ S2 in the
Supplementary Material).  Here we have chosen to use $n = 5$
(equivalently $T = 5$ d) as this guarantees both good interbox
communication and negligible memory into the past.  Similar choices
have been made in recent applications involving drifter data
\cite{Maximenko-etal-12, Miron-etal-17, McAdam-vanSebille-18,
Miron-etal-18a, Olascoaga-etal-18}.

The simplest choice for the time-homogeneity interval $\mathfrak
T$ is one that coincides with the entire record of trajectory data
\cite{Maximenko-etal-12, Miron-etal-17, McAdam-vanSebille-18,
Miron-etal-18a, Olascoaga-etal-18}.  The resulting \emph{autonomous}
Markov chain, which will be only considered for comparison purposes,
ignores any mode of variability of the ocean circulation and thus
is not optimal for describing debris motion in seasonally dependent
environments like the Indian Ocean.

Different $\mathfrak T$ intervals can be considered (e.g., van
Sebille et al. \cite{vanSebille-etal-12}) to represent the dominant
variability mode of the Indian Ocean circulation, produced by
seasonal changes in the wind stress associated with the Indian
monsoon \cite{Schott-McCreary-01}. During the northern winter,
when the monsoon blows southwestward, the flow of the upper ocean
is directed westward from near the Indonesian Archipelago to the
Arabian Sea.  During the northern summer, with the change of the
monsoon direction toward the northeast, the ocean circulation
reverses, with eastward flow extending from Somalia into the Bay
of Bengal.  Thus we consider three $\mathfrak T$ intervals:
January--March ($\mathfrak T_\mathrm{W}$), which typically corresponds
to the winter monsoon season, July--September ($\mathfrak T_\mathrm{S}$),
corresponding to the summer monsoon season, and April--June and
October--December together ($\mathfrak T_\mathrm{SF}$), seasons
which do not need to be distinguished from one another to represent
the monsoon-induced circulation of the Indian Ocean.  This results
in three transition matrices, $P_\mathrm{W}$, $P_\mathrm{S}$ and
$P_\mathrm{SF}$, respectively, which are appropriately considered
for $t\in \mathfrak T_\mathrm{W}$, $\mathfrak T_\mathrm{S}$ or $
\mathfrak T_\mathrm{SF}$, when a probability vector is evolved
(pushed forward) under left multiplication.  We will refer to the
resulting Markov-chain model as \emph{nonautonomous}.

Finally, if the interest is in the fate of the debris in the
seasonally changing Indian Ocean environment after several years,
one can more conveniently push forward probabilities using a $P$
constructed by combining the above seasonal $P$s in such a way that
the resulting Markov chain has a transition time $T$ of 1 yr.
Recalling that $T = 5$ d for the seasonal $P$s, this is (approximately)
achieved by $P = P_\mathrm{W}^{18}\cdot P_\mathrm{SF}^{18}\cdot
P_\mathrm{S}^{18}\cdot P_\mathrm{SF}^{18}$.  The resulting Markov-chain
model will be referred to as \emph{autonomous season-aware}.  Similar
constructions have been considered earlier (e.g., Froyland et al.\
\cite{Froyland-etal-14}).

\begin{figure*}[t!]
  \centering
  \includegraphics[width=\textwidth]{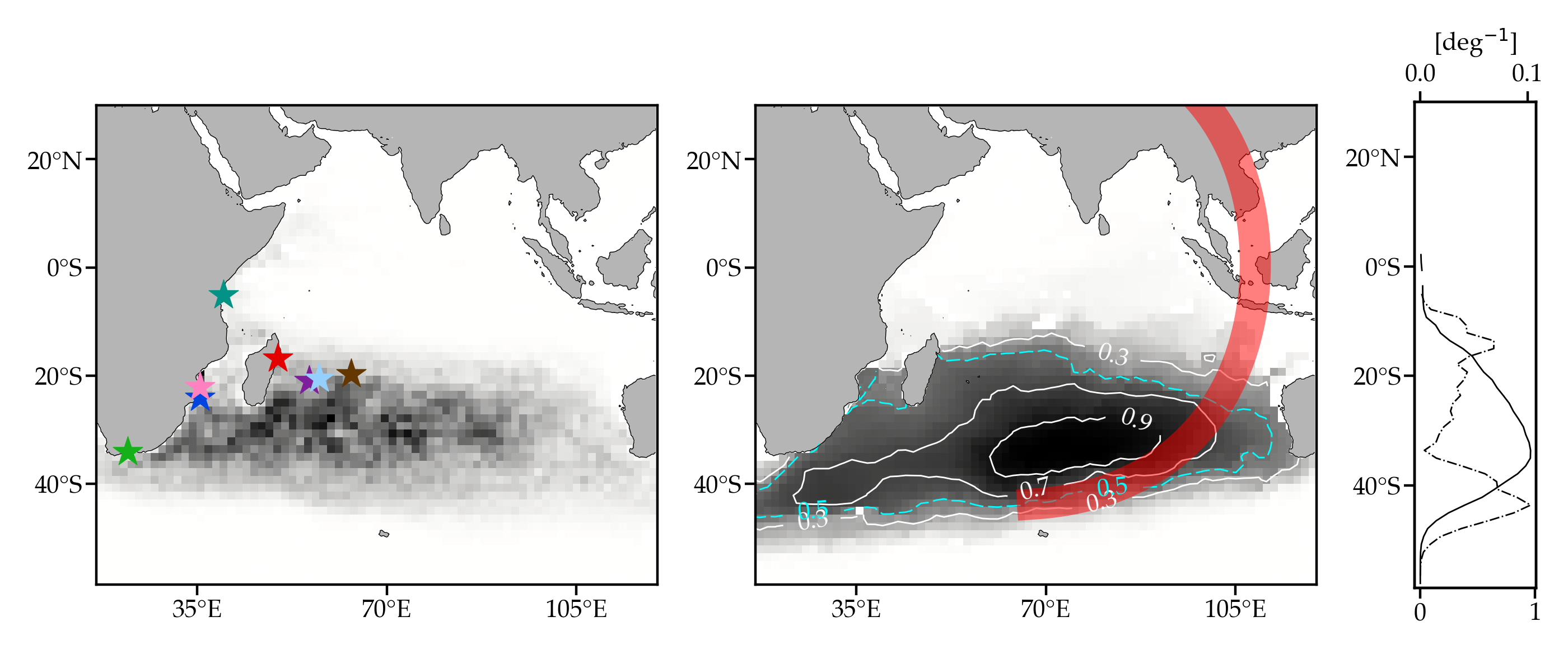}%
  \caption{(left) Dominant left eigenvector of the autonomous
  season-aware transition matrix with beaching sites indicated (cf.\
  Table \ref{tab:debris}).  (middle) Dominant right eigenvector
  with the Inmarsat arc overlaid.  (right) Zonally averaged right
  eigenvector (solid, bottom axis) and its derivative (dashed, top
  axis).  Local maxima in the left eigenvector are regions that
  attract trajectories which tend to run long there before exiting
  the Indian Ocean.  The basin of attraction roughly corresponds
  to the region enclosed by the 0.5 right eigenvector level set,
  above which the eigenvector looks closer to $\mathbf 1$.}
  \label{fig:geo}%
\end{figure*}

\section{Crash site assessment from spectral analysis}

Information about the long-time asymptotic behavior of a dynamical
system described by an autonomous transition matrix $P$ can be
obtained from its spectral properties \cite{Hsu-87, Dellnitz-Junge-99,
Froyland-05}. Indeed, the leading eigenvector structure of $P$
suggests a dynamical partition or \emph{geography} \cite{Froyland-etal-14,
Miron-etal-17, Miron-etal-18a} of weakly interacting sets that
constrains connectivity between distant points in phase space.  Here
we unveil such a dynamical geography from the autonomous season-aware
Markov chain to gain insight into debris motion and thus the crash
site.

For any irreducible and aperiodic stochastic $P$, its dominant
\emph{left} eigenvector, $\mathbf p$, satisfies $\mathbf p P =
\mathbf p$ and (scaled appropriately) represents a limiting invariant
or \emph{stationary} distribution, namely, $\mathbf{p} =
\smash{\lim_{k\uparrow \infty} \mathbf f P^k}$ for any probability
vector $\mathbf f$ (cf., e.g., Horn and Johnson \cite{Horn-Johnson-90}).
Also, $\mathbf 1 = P\mathbf 1$, where $\mathbf 1$ denotes the vectors
of ones, is the \emph{right} eigenvector corresponding to the
eigenvalue 1 and $\mathbf p$.

If $P$ is substochastic, i.e., $\smash{\sum_j} P_{ij} < 1$ for some
$i$, the dominant left eigenvector $\mathbf{p} \ge 0$ decays at a
rate set by the dominant eigenvalue $\lambda_1 < 1$, and has the
interpretation of a limiting almost-invariant or \emph{quasistationary}
distribution, namely, the limiting distribution of trajectories
that run long before terminating (cf.\ Chapter 6.1.2 of Bremaud
\cite{Bremaud-99}).  Restricted to the set $B$ where those trajectories
start, i.e., the \emph{basin of attraction}, the dominant right
eigenvector is close to $\mathbf 1$ \cite{Koltai-11, Froyland-etal-14,
Miron-etal-17, Miron-etal-18a}.

The left and middle panels of Fig.\ \ref{fig:geo} show respectively
the dominant left ($\mathbf p$) and right ($\mathbf r$) eigenvectors
of the autonomous $P$ introduced in the preceding section.  The
right panel in turn shows again the right eigenvector, but this
time zonally averaged (solid, bottom axis), along with its (meridional)
derivative (dashed, top axis), which maximizes near the 0.5 level
set.  The dominant eigenvalue $\lambda_1 = 0.8181$ sets an annual
decay rate for $\mathbf p$ of about 20\pct, which is nearly four
times slower than that experienced by the first subdominant left
eigenvector ($\lambda_2 = 0.4012$). Note the structure of $\mathbf
p$, taking many local maxima toward the western side of Indian
Ocean, inside the 20--40$^{\circ}$S band.  The right eigenvector
is much less structured, looking closer to $\mathbf 1$ within the
region enclosed by the 0.5 level set.  This region approximately
forms a basin of attraction $B = \{\mathbf r > 0.5\}$ for trajectories
that asymptotically distribute as $\mathbf p$ conditional to staying
in the domain for a long time.

The expected retention time in $B$ is given by $T_B = T/(1 -
\lambda_B)$, where $\lambda_B$ is the dominant eigenvalue of $P$
restricted to $B$ (cf.\ Appendix B of the Supplementary Material
for a derivation and Miron et al.\ \cite{Miron-etal-18a} for a
recent application).  We compute $\lambda_B = 0.5103$, and noting
that $T = 1$ yr for the autonomous season-aware $P$, we estimate
$T_B = 2.0421$ yr, which is of the order of the mean time it took
observed airplane debris to reach the African coasts (cf.\ Table
\ref{tab:debris}).  This long residence time and the large area
spanned by $B$ impose a constraint on the connectivity between
locations in- and outside of $B$ by debris trajectories, which we
use to make an initial assessment of the possible crash site as
follows.

Note in the left panel of Fig.\ \ref{fig:geo} that the observed
beaching sites (stars) lie within or the border of the western
region where $\mathbf p$ tends to locally maximize.  Note in the
middle panel that the Inmarsat arc traverses the eastern side of
$B$.  This suggests a possible crash site somewhere along the
Inmarsat arc sector between the latitudes of intersection with of
$B$, roughly 20 and 40$^{\circ}$S.

In the next sections we will show that the uncertainty (about 3600
km) of the spectral assessment above can be substantially reduced
using a dedicated Bayesian analysis along with the computation of
most probable paths and the inspection of a particular drifter
trajectory.

\section{Bayesian estimation of the crash site}

The Bayesian analysis uses the beaching events as observations to
infer the probability distribution of the crash site (refer to
Appendices C and D in the Supplementary Material for mathematical
details).  Because of the shorter-time nature of these observations,
the analysis is most appropriately carried out using the nonautonomous
Markov-chain model, albeit with some convenient adaptation.

As our computational domain is not the whole ocean, necessarily
trajectories can leave the domain, and not have an assignable
termination box.  For example, the Agulhas Current transports
approximately 70 Sv (1 Sv $= 10^6$ m$^3$s$^{-1}$) out of the western
boundary of the domain \cite{Beal-etal-11}.  This leakage is
represented by $\smash{\sum_j} P_{ij} < 1$ for the leaky states
$i\in\mathfrak L\subset S := \{1, \dotsc, N\}$ (black boxes in Fig.\
\ref{fig:grid}).  (Herein $P$ is any $P_\mathrm{W}$, $P_\mathrm{S}$
or $P_\mathrm{SF}$, which are applied for $t\in \mathfrak T_\mathrm{W}$,
$\mathfrak T_\mathrm{S}$ or $ \mathfrak T_\mathrm{SF}$ as appropriate.)
To account for the leakage and retain the open dynamics nature of
the Indian Ocean, we introduce an absorbing state $N+1$, commonly
referred to as \emph{cemetery} state, and consider $1 - \smash{\sum_j}
P_{i\in\mathfrak L,j}$ to be the probability of the chain to move
and terminate in it, if currently being in state $i\in \mathfrak
L$.  This amounts to augmenting $P\in\mathbb R^{N\times N}$ to
$P\in\mathbb R^{(N+1)\times (N+1)}$ by
\begin{equation}
  \begin{cases} 
	 P_{i\in \mathfrak L,j=N+1} \leftarrow 1 - \sum_j P_{i\in \mathfrak
	 L,j},\\
	 P_{i=N+1,i}  = 1, 
  \end{cases}
\end{equation}
satisfying $\sum_j P_{ij} = 1$.

In addition to the dynamics being open, as we are considering
floating debris which might come ashore somewhere, the transition
matrix has to account for the Markov chain possibly terminating on
water--land interfaces. Such events are hard to identify from the
dataset because drifters can terminate for multiple reasons (e.g.,
malfunction, recovery, end of life).  To account for beaching, we
identify the set of sticky states $\mathfrak S \subset S$ (grey
boxes in Fig.\ \ref{fig:grid}) with those coastal boxes that contain
land--water interfaces and introduce the \emph{land fraction}
function $\ell : \mathfrak S \to (0,1)$.  Note that a sticky state
may also be leaky, namely, $\mathfrak S \cap \mathfrak L$ need not
be empty. We then denote by $\mathfrak D \subset \mathfrak S$ the
subset of sticky boxes where airplane debris were found to beach,
to distinguish them from the others.  Beaching is then modeled by
augmenting $P\in\smash{\mathbb R^{(N+1)\times (N+1)}}$ to
$P\in\smash{\mathbb R^{(N+1+M)\times (N+1+M)}}$, satisfying $\sum_j
P_{ij} = 1$ and where $M = |\mathfrak D| (=8)$, according to:
\begin{equation}
  \begin{cases}
  P_{i\in \mathfrak S,j\in S\cup \{N+1\}} \leftarrow
  (1\!-\!\ell(i\!\in\!\mathfrak S)) P_{i\in \mathfrak S,j
  \in S\cup \{N+1\}},\\ P_{i\in \bar{\mathfrak S},j=N+1} \leftarrow
  \ell(i\!\in\!\bar{\mathfrak S}) + (1\!-\!\ell(i\!\in\!\bar{\mathfrak
  S})) P_{i\in \bar{\mathfrak S},j=N+1},\\ P_{i\in \mathfrak
  D,j=N+1+m(i\in \mathfrak D)} = \ell(i\!\in\!\mathfrak D),\\
  P_{i \in N+1+m(\mathfrak D),i} = 1,
  \end{cases}
  \label{eq:Pa}
\end{equation}
where $\bar{\mathfrak S} := \mathfrak S \setminus \mathfrak D$, $m
: \mathfrak D \to \{1, \dotsc, M\}$, and $N+1+m(i\in \mathfrak D)$
represents a \emph{target cemetery} state where the chain terminates
whenever beaching from $i\in \mathfrak D$. The first line in
\eqref{eq:Pa} deals with sticky boxes if the chain flows on.  The
second line deals with sticky but nondebris beaching boxes, each
of which also is terminated in the cemetery state ($N+1$). The third
line deals with debris-beaching boxes if the chain terminates from
them by beaching.  Finally, the fourth line makes each target
cemetery state absorbing.

\begin{figure}[t]
  \centering%
  \includegraphics[width=\columnwidth]{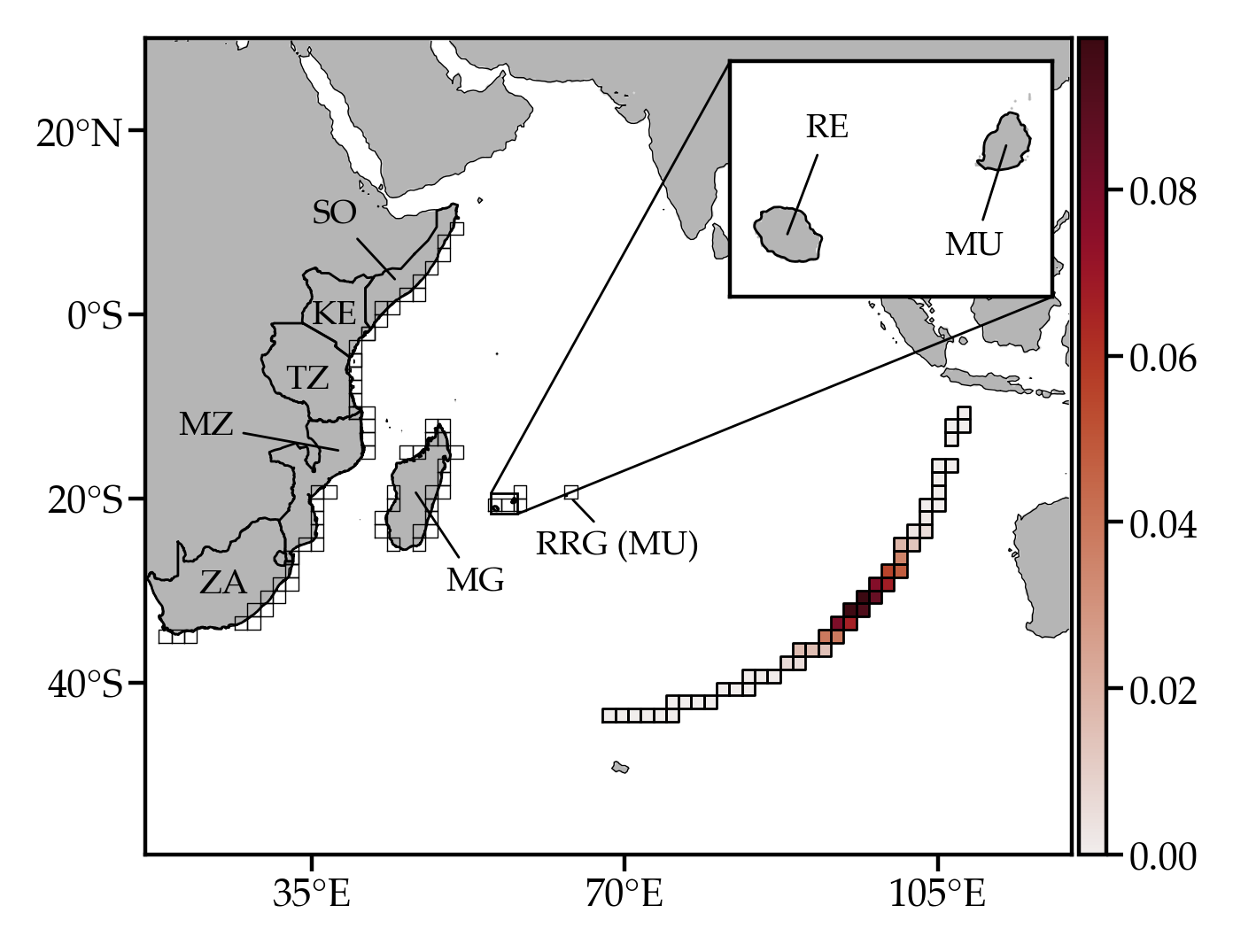}% 
  \caption{Posterior probability of the crash site, for given assumed
  mutually independent observations of debris beaching time and
  based on the nonautonmous Markov-chain model.  Indicated are
  countries and territories in the western side of the Indian Ocean
  domain and corresponding boxes of the domain covering.  Acronyms
  SO and KE stand for Somalia and Kenya, respectively; cf.\ Table
  \ref{tab:debris} for the rest of the acronyms.}
  \label{fig:post}%
\end{figure}

Having settled on an appropriate $P$ representation, we proceed to
formulate the Bayesian estimation problem by first fixing some
notation.  Let $\mathfrak C \subset S$ denote the set of indices
of the boxes (states) along the Inmarsat arc---the possible crash
sites.  We call $b$ a state in the set $\mathfrak B := \{N+2, \dotsc,
N+M+1\}$ of airplane debris beaching site boxes and further make
$\mathbf b := (b)_{b\in\mathfrak B}\in\mathbb{N}^M$. Let
$\{\xi_{t+kT}\}_{k\ge 0}$ be time-discrete random position variables
that take values on the augmented Markov chain states $S \cup \{N+1\}
\cup \mathfrak B$, and $t$ denotes the crash time, when the chain
starts.  Consider then the random variable $\tau^b$ denoting the
time after which the chain gets absorbed into a particular beaching
state $b$, namely,
\begin{equation}
  \tau^b := \inf_{k\ge 0}\{t+kT : \xi_{t+kT} = b\in \mathfrak B\}.
  \label{eq:tau}
\end{equation}
Define $\Pr_c[\cdot] :=  \Pr[\,\cdot \mid \xi_t = c\in \mathfrak C]$ and
\begin{equation}
  p_c^b(k) := \mathbf 1_c P^k \cdot \mathbf 1_b,\quad c\in
  \mathfrak C,\, b\in \mathfrak B,
\end{equation}
where $\mathbf 1_j = (\delta_{ij})_{i\in S \cup
\{N+1\} \cup \mathfrak B}$, and then note that
\begin{equation}
  \Pr_c[\tau^b = t+kT] = 
  \begin{cases} 
   p_c^b(k) = 0           & \text{if } k = 0,\\
   p_c^b(k) - p_c^b(k-1)  & \text{if } k > 0.
  \end{cases}
  \label{eq:p}
\end{equation}

Let us now assume that after the crash every single piece of debris
was transported mutually independently. For each debris, we have
access to two observations of random quantities: the \emph{beaching
site} and the \emph{beaching time}. Let $\xi$ and $\tau$ denote
these random variables, respectively, and note that if the target
is absorbing, then $\tau^b < \infty$ from \eqref{eq:tau} implies
$\xi = b$, and thus the events $\{\xi = b \text{ and } \tau = t^b\}$
and $\{\tau^b = t^b\}$ are equivalent.  Thus the joint probability
of these observations, i.e.,
\begin{equation}
  p(t^b|c) := \Pr_c[\tau = t^b \text{ and } \xi = b]
  \equiv \Pr_c[\tau^b = t^b],
\end{equation}
is computable as in \eqref{eq:p}.  The probabilities $p(t^b|c)$
depend on the crash site $c$. Now, making use of the independence
assumption above, the joint probability of the observations, given
the crash site $c$, is
\begin{equation}
  p(t^\mathbf{b}|c) := \prod_{b\in \mathfrak B} p(t^b|c). 
\end{equation}

The idea of Bayesian inversion \cite{Bolstad-Curran-16} is to find
a probabilistic characterization of the unknown parameter $c$ (crash
site) given the debris beaching times and locations.  By viewing
$p(t^\mathbf{b}|c)$ as a function of $c$, one obtains a function
$L(c;t^\mathbf{b})$ which represents the \emph{likelihood} of $c$.
The \emph{posterior distribution} of $c$, i.e., the probability
distribution of $c$ once $t^\mathbf{b}$ have been observed, follows
from Bayes' theorem:
\begin{equation}
  p(c|t^\mathbf{b}) \propto p(t^\mathbf{b}|c)\cdot p(c),
  \label{eq:post}
\end{equation}
where $p(c)$ is the \emph{prior distribution} of $c$, representing
the state of knowledge about $c$ before to data have been observed.

Assuming that $p(c)$ is uniform (i.e., maximally uninformative)
over $\mathfrak C$, the Inmarsat arc boxes, Fig.\ \ref{fig:post}
shows $p(c|t^\mathbf{b})$ for debris beaching times $t^\mathbf{b}$
as given in Table \ref{tab:debris}.  The \emph{maximum likelihood
estimator}, $c_\mathrm{max} = \argmax_c p(t^\mathbf{b}|c)$, corresponds
to the index of the Inmarsat arc box centered at about %(96.6$^{\circ}$E, 30.7$^{\circ}$S).
31$^{\circ}$S. To check how reasonable this crash site estimate is,
the top panel of Fig.\ S3 in Appendix F of the Supplementary Material
shows that the joint probability $\Pr_{c_\mathrm{max}}[\tau = t+kT
\text{ and } \xi_{t+kT} = s]$ for each sticky state $s\in \mathfrak
S$ tends to maximize near the observations.  The misfit is attributed
to the historical drifter data not capturing all of the details of
the Indian Ocean dynamics and in particular those when the crash
took place and the years after it.  For instance, the bottom panel
of Fig.\ S3 shows that the misfit augments when the monsoon variability
is ignored.

\begin{figure}[t]
  \centering%
  \includegraphics[width=\columnwidth]{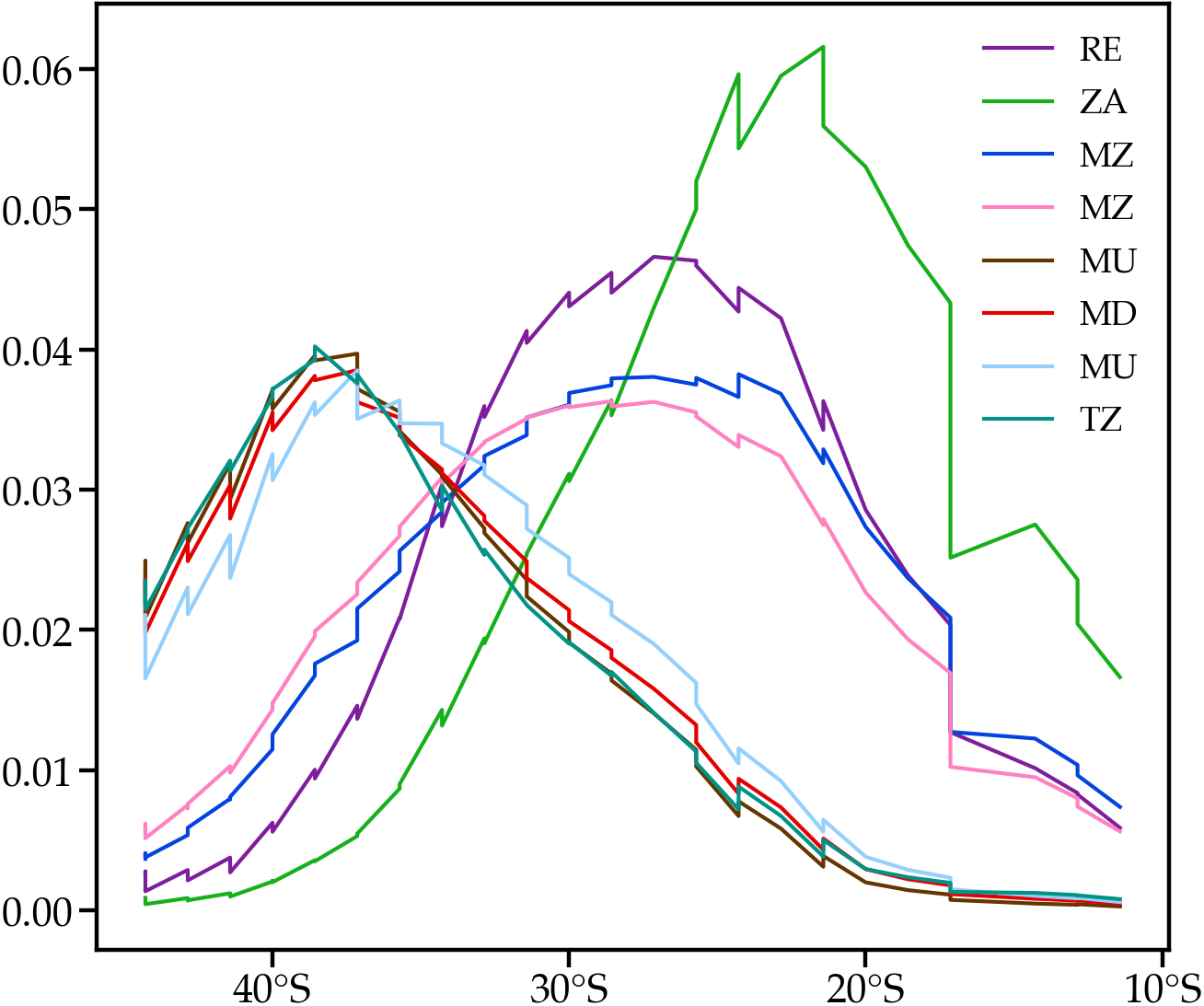}%
  \caption{Single posterior distributions of the latitude of the
  crash site along the Inmarsat arc, computed using the nonautonmous
  Markov-chain model. Refer to Table \ref{tab:debris} for
  acronym meanings.}
  \label{fig:Lb}%
\end{figure}

The Bayesian crash site estimate, %(96.6$^{\circ}$E, 30.7$^{\circ}$S),
31$^{\circ}$S on the Immarsat arc, lies within the arc portion
identified as a likely crash region using the spectral analysis of
the previous section.  An important difference is that the uncertainty
of the assessment is constrained by the Bayesian analysis.  For
instance, the 95\pct\, central posterior interval length, obtained
by computing the 2.5 and 97.5\pct-ile of the posterior distribution,
is of about %1700 km along the arc, which is nearly 2.5 times narrower than 
12$^{\circ}$ along the arc, which is nearly twice as narrow as
the spectral inference. However, %for practical purposes, this
%uncertainty might still be considered too large, and is explained
the Bayessian inference is affected
by a bimodality in the single posterior distributions of the crash
site.  This is demonstrated in Fig.\ \ref{fig:Lb}, which shows
$p(c\vert t^b)$ plotted for each $b\in\mathfrak B$ individually as
a function of the latitude of $c \in \mathfrak C$. These can be
collected into two quite consistent groups.  Specifically, a southern
group, which favors a most likely crash site near $36^{\circ}$S on
the arc, and a northern group, favoring a most likely position near
$25^{\circ}$S.

\section{Narrowing the uncertainty of the crash site determination}

We close the analysis by discussing one additional calculation and
a particular observation that altogether provide means for favoring
the northern of the above two possible crash sites, %narrowing the
%uncertainty around its determination to about 750 km.
improving the confidence of its determination.

Drifter data does not merely give the chance to construct a Markov
chain model of the surface currents, but also to study single
trajectories. Thus we consider the most probable paths of our model
that end up at the particular beaching sites. More precisely, as
we know how long (in time) the single trajectories were, we compute
most probable paths ending at a given beaching site $b\in \mathfrak
B$ after having ran for a fixed time $t^b$.  The basic idea of the
process is to set up and iteratively solve a dynamical programming
equation relating the maximal probability of a path reaching some
state $i\in S$ after exactly $t + T$ units of time with the maximal
probabilities of paths reaching other states $j\in S$ after $t$
(cf.\ Appendix E in the Supplementary Material for details as well
as for a review of standard unconstrained extremal path notions
\cite{Dijkstra-59, Floyd-62}).  It is important to realize that the
probability of \emph{single} paths is extremely low (as there are
so many possible ones), and the most probable one can sensitively
depend on the slightest variations of the dynamics.  Thus they are
only suitable for qualitative comparisons with observations.  We
note that constrained paths were considered to analyze currents in
the Mediterranean Sea by Ser-Giacomi et al.\ \cite{SerGiacomi-etal-15b},
yet without excluding the possibility to hit the target before the
end time.

\begin{figure}[t]
  \centering%
  \includegraphics[width=\columnwidth]{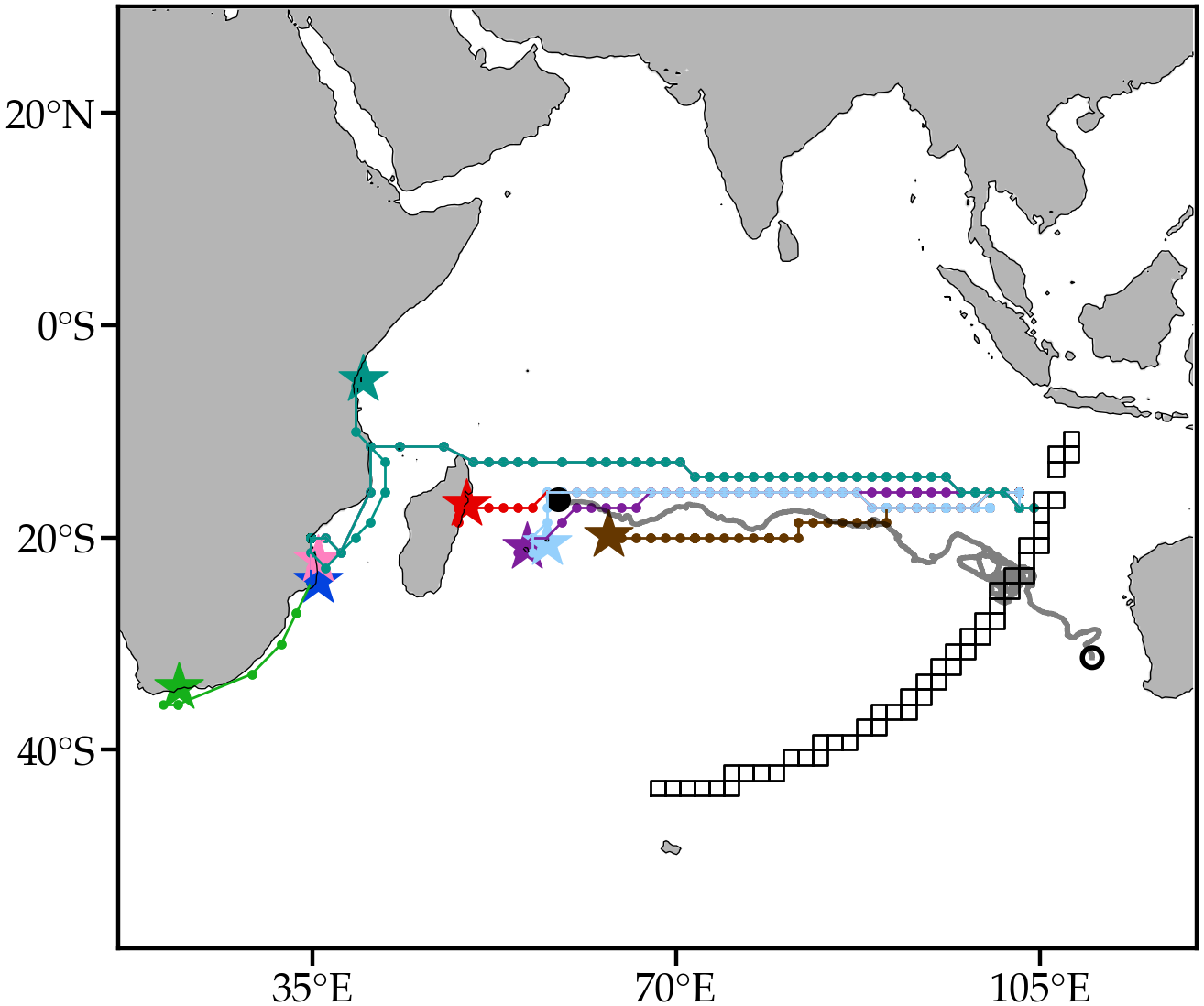}%
  \caption{Most probable paths (dotted) along the nonautonomous
  Markov chain between Inmarsat arc boxes and each of the debris
  beaching sites (cf.\ Table \ref{tab:debris}), and trajectory
  (gray) of a NOOA/GDP drifter (ID 56568) launched west of the
  Australian coast.}
  \label{fig:mp_path}%
\end{figure}

The result is shown in Fig.~\ref{fig:mp_path}, revealing, on the
one hand, quite remarkably that the most probable paths of different
fixed lengths (dotted curves) start from a common box that is the
Inmarsat arc box centered at roughly (105$^{\circ}$E, 16$^{\circ}$S).
On the other hand, Tri\~nanes et al.\ \cite{Trinanes-etal-16} report
on a drifter (ID 56568, from the NOAA/GDP database) that crossed
the Inmarsat arc near (103$^{\circ}$E, 22$^{\circ}$S) in March 2014
and after looping for a little while about the arc near (102$^{\circ}$E,
25$^{\circ}$S) reached the vicinity of Reunion Island in July 2015,
about the time when the first MH370 debris piece was spotted.  The
trajectory of the drifter in question is shown in gray in
Fig.~\ref{fig:mp_path}.

Taken together, the most probable path computation and the specific
drifter trajectory observation favor the Bayesian crash site estimate
near 25$^{\circ}$.  Excluding from the Bayesian analysis the debris
beaching data that leads to single posterior crash site distributions
peaking near $36^{\circ}$S, we estimate a 95\pct\, central posterior
interval length of 16$^{\circ}$ along the Inmarsat arc for the refined
crash site inference.

We finally note that the qualitative assessment in this section,
along with the spectral analysis assessment, provide a guideline
to construct a prior distribution that can be incorporated to the
Bayesian analysis. More specifically, these assessments suggest an
informative prior distribution narrowing near 25$^{\circ}$N on the
Inmarsat arc.

\section{Summary and concluding remarks}

Using historical satellite-tracked surface drifter data in the
Indian Ocean, we have proposed a Markov-chain model representation
of the drift of observed marine debris from the missing Malaysian
Airlines flight MH370. 

The results from a spectral analysis of an autonomous discrete
transfer operator (transition matrix) that controls the long-term
evolution of the debris in a seasonally changing environment showed
that the crash region is likely restricted to the 20--40$^{\circ}$S
portion of the arc of the 7th ping ring around the Inmarsat-3F1
satellite position when the airplane lost contact on 8 March 2014.
The solution of a dedicated Bayesian estimation problem that uses
the locations and times of confirmed airplane debris beachings in
a Markov-chain model defined by a nonutonomous transition matrix
capable of resolving shorter-term details of the debris evolution,
further identified two probable crash sites within the aforementioned
arc portion, one near $36^{\circ}$S and another one near $25^{\circ}$S.
Consideration of most probable paths between the Inmarsat arc and
the debris beaching sites constrained by the observed (elapsed)
beaching times, and the observation of a drifter that took a
trajectory similar to the path connecting the arc and the Reunion
Island beaching site, was shown when taken together to favor the
25$^{\circ}$S crash site estimate.

Our Bayesian crash site estimate, with a 95\pct\, central posterior
interval ranging from 33 to 17$^{\circ}$S on the Inmarsat arc, is
consistent with the most recently published \cite{Nesterov-18} crash
area estimate, 28--30$^{\circ}$ along the arc, but it lies north
of the latest recommended search area by the Commonwealth Scientific
and Industrial Research Organisation \cite{Griffin-etal-17}, at
around 35$^{\circ}$S.  Notwithstanding, it is consistent with the
original northern definition of high-priority search zone by the
Australian Transport Safety Bureau \cite{ATSB-14}.

The uncertainty of our crash site estimate may be attributed to
unresolved nonautonomous dynamical effects by the Markov-chain model
constructed from the historical drifter data. Indeed, the historical
drifter data may fall short of capturing all of the details of the
circulation (e.g., at the submesoscale), particularly during the
years after the crash took place.  Moreover, undrogued drifter
motion, while different than water parcel motion, may not accurately
represent the motion of airplane debris pieces with varied shapes
and thus drag properties.

While our results may provide grounds for guiding search efforts,
currently halted, the operational use of the probabilistic framework
proposed here in such a task will require one to consider an
appropriate data-assimilative system.  Indeed, the probabilistic
framework may well be applied on observed trajectory data, as we
have chosen to do in this paper, or on numerically generated
trajectory data.  Clearly, the success of such an operative use of
the framework will depend on the operator's ability to appropriately
modeling the effects of the inertia of the debris pieces (i.e., of
their buoyancy, size and shape), which is a subject of active
research \cite{Cartwright-etal-10, Beron-etal-15, Beron-etal-16}.

We finally note that the framework here presented is well-suited
for inverse modeling in a general setting and thus its utility is
far reaching.  Such modeling is critical, for instance, in contamination
source backtracking \cite{Bagtzoglou-Atmadja-05}. Relevant oceanic
problems include that of red tide early development tracing
\cite{Olascoaga-etal-08} and oil spill source detection
\cite{Gautama-etal-16}. In the atmosphere, examples for instance
are the identification of sources of greenhouse gases emission
\cite{Hourdin-Talagrand-06} and toxic agents release
\cite{Rao-07}.

\begin{acknowledgments}
The comments by an anonymous reviewer help improved the paper. We
thank Ilja Klebanov and Tim Sullivan for the benefit of discussions
on Bayesian inference.  The NOAA/GDP dataset is available at
http://\allowbreak www.aoml.noaa.gov/\allowbreak phod/\allowbreak
dac/. Some techniques employed in this paper were discussed by FJBV,
MJO and PK during the Erwin Schr\"odinger Institute (ESI) Programme
on Mathematical Aspects of Physical Oceanography. Support from ESI
is greatly appreciated. The work of PK was partially supported by
the Deutsche Forschungsgemeinschaft (DFG) through the Priority
Programme SPP 1881 Turbulent Superstructures and through the CRC
1114 Scaling Cascades in Complex Systems project A01.
\end{acknowledgments}

%\bibliography{fot,Literature}

\begin{thebibliography}{52}%
\makeatletter
\providecommand \@ifxundefined [1]{%
 \@ifx{#1\undefined}
}%
\providecommand \@ifnum [1]{%
 \ifnum #1\expandafter \@firstoftwo
 \else \expandafter \@secondoftwo
 \fi
}%
\providecommand \@ifx [1]{%
 \ifx #1\expandafter \@firstoftwo
 \else \expandafter \@secondoftwo
 \fi
}%
\providecommand \natexlab [1]{#1}%
\providecommand \enquote  [1]{``#1''}%
\providecommand \bibnamefont  [1]{#1}%
\providecommand \bibfnamefont [1]{#1}%
\providecommand \citenamefont [1]{#1}%
\providecommand \href@noop [0]{\@secondoftwo}%
\providecommand \href [0]{\begingroup \@sanitize@url \@href}%
\providecommand \@href[1]{\@@startlink{#1}\@@href}%
\providecommand \@@href[1]{\endgroup#1\@@endlink}%
\providecommand \@sanitize@url [0]{\catcode `\\12\catcode `\$12\catcode
  `\&12\catcode `\#12\catcode `\^12\catcode `\_12\catcode `\%12\relax}%
\providecommand \@@startlink[1]{}%
\providecommand \@@endlink[0]{}%
\providecommand \url  [0]{\begingroup\@sanitize@url \@url }%
\providecommand \@url [1]{\endgroup\@href {#1}{\urlprefix }}%
\providecommand \urlprefix  [0]{URL }%
\providecommand \Eprint [0]{\href }%
\providecommand \doibase [0]{http://dx.doi.org/}%
\providecommand \selectlanguage [0]{\@gobble}%
\providecommand \bibinfo  [0]{\@secondoftwo}%
\providecommand \bibfield  [0]{\@secondoftwo}%
\providecommand \translation [1]{[#1]}%
\providecommand \BibitemOpen [0]{}%
\providecommand \bibitemStop [0]{}%
\providecommand \bibitemNoStop [0]{.\EOS\space}%
\providecommand \EOS [0]{\spacefactor3000\relax}%
\providecommand \BibitemShut  [1]{\csname bibitem#1\endcsname}%
\let\auto@bib@innerbib\@empty
%</preamble>
\bibitem [{\citenamefont {Ashton}\ \emph {et~al.}(2015)\citenamefont {Ashton},
  \citenamefont {Bruce}, \citenamefont {Colledge},\ and\ \citenamefont
  {Dickinson}}]{Ashton-etal-15}%
  \BibitemOpen
  \bibfield  {author} {\bibinfo {author} {\bibfnamefont {C.}~\bibnamefont
  {Ashton}}, \bibinfo {author} {\bibfnamefont {A.~S.}\ \bibnamefont {Bruce}},
  \bibinfo {author} {\bibfnamefont {G.}~\bibnamefont {Colledge}}, \ and\
  \bibinfo {author} {\bibfnamefont {M.}~\bibnamefont {Dickinson}},\ }\bibfield
  {title} {\enquote {\bibinfo {title} {{The search for MH370}},}\ }\href@noop
  {} {\bibfield  {journal} {\bibinfo  {journal} {The Journal of Navigation}\
  }\textbf {\bibinfo {volume} {68}},\ \bibinfo {pages} {1--22} (\bibinfo {year}
  {2015})}\BibitemShut {NoStop}%
\bibitem [{\citenamefont {Holland}(2018)}]{Holland-18}%
  \BibitemOpen
  \bibfield  {author} {\bibinfo {author} {\bibfnamefont {I.~D.}\ \bibnamefont
  {Holland}},\ }\bibfield  {title} {\enquote {\bibinfo {title} {{MH370} burst
  frequency offset analysis and implications on descent rate at end of
  flight},}\ }\href@noop {} {\bibfield  {journal} {\bibinfo  {journal} {IEEE
  Aerospace and Electronic Systems Magazine}\ }\textbf {\bibinfo {volume}
  {33}},\ \bibinfo {pages} {24--33} (\bibinfo {year} {2018})}\BibitemShut
  {NoStop}%
\bibitem [{\citenamefont {{Australian Transport Safety
  Bureau}}(2018)}]{ATBS-18}%
  \BibitemOpen
  \bibfield  {author} {\bibinfo {author} {\bibnamefont {{Australian Transport
  Safety Bureau}}},\ }\href
  {https://www.atsb.gov.au/publications/investigation\_reports/2014/aair/ae-2014-054/}
  {\enquote {\bibinfo {title} {{Assistance to Malaysian Ministry of Transport
  in support of missing Malaysia Airlines flight MH370 on 7 March 2014 UTC}},}\
  }\bibinfo {howpublished} {Investigation number: AE-2014-054} (\bibinfo {year}
  {2018})\BibitemShut {NoStop}%
\bibitem [{\citenamefont {Garc\'{\i}a-Garrido}\ \emph
  {et~al.}(2015)\citenamefont {Garc\'{\i}a-Garrido}, \citenamefont {Mancho},
  \citenamefont {Wiggins},\ and\ \citenamefont {Mendoza}}]{Garcia-etal-15}%
  \BibitemOpen
  \bibfield  {author} {\bibinfo {author} {\bibfnamefont {V.~J.}\ \bibnamefont
  {Garc\'{\i}a-Garrido}}, \bibinfo {author} {\bibfnamefont {A.~M.}\
  \bibnamefont {Mancho}}, \bibinfo {author} {\bibfnamefont {S.}~\bibnamefont
  {Wiggins}}, \ and\ \bibinfo {author} {\bibfnamefont {C.}~\bibnamefont
  {Mendoza}},\ }\bibfield  {title} {\enquote {\bibinfo {title} {{A dynamical
  systems approach to the surface search for debris associated with the
  disappearance of flight MH370}},}\ }\href@noop {} {\bibfield  {journal}
  {\bibinfo  {journal} {Nonlinear Processes in Geophysics}\ }\textbf {\bibinfo
  {volume} {22}},\ \bibinfo {pages} {701--712} (\bibinfo {year}
  {2015})}\BibitemShut {NoStop}%
\bibitem [{\citenamefont {Corrado}\ \emph {et~al.}(2017)\citenamefont
  {Corrado}, \citenamefont {Lacorata}, \citenamefont {Palatella}, \citenamefont
  {Santoleri},\ and\ \citenamefont {Zambianchi}}]{Corrado-etal-17}%
  \BibitemOpen
  \bibfield  {author} {\bibinfo {author} {\bibfnamefont {R.}~\bibnamefont
  {Corrado}}, \bibinfo {author} {\bibfnamefont {G.}~\bibnamefont {Lacorata}},
  \bibinfo {author} {\bibfnamefont {L.}~\bibnamefont {Palatella}}, \bibinfo
  {author} {\bibfnamefont {R.}~\bibnamefont {Santoleri}}, \ and\ \bibinfo
  {author} {\bibfnamefont {E.}~\bibnamefont {Zambianchi}},\ }\bibfield  {title}
  {\enquote {\bibinfo {title} {General characteristics of relative dispersion
  in the ocean},}\ }\href@noop {} {\bibfield  {journal} {\bibinfo  {journal}
  {Scientific Reports}\ }\textbf {\bibinfo {volume} {7}},\ \bibinfo {pages}
  {46291} (\bibinfo {year} {2017})}\BibitemShut {NoStop}%
\bibitem [{\citenamefont {Trinanes}\ \emph {et~al.}(2016)\citenamefont
  {Trinanes}, \citenamefont {Olascoaga}, \citenamefont {Goni}, \citenamefont
  {Maximenko}, \citenamefont {Griffin},\ and\ \citenamefont
  {Hafner}}]{Trinanes-etal-16}%
  \BibitemOpen
  \bibfield  {author} {\bibinfo {author} {\bibfnamefont {J.~A.}\ \bibnamefont
  {Trinanes}}, \bibinfo {author} {\bibfnamefont {M.~J.}\ \bibnamefont
  {Olascoaga}}, \bibinfo {author} {\bibfnamefont {G.~J.}\ \bibnamefont {Goni}},
  \bibinfo {author} {\bibfnamefont {N.~A.}\ \bibnamefont {Maximenko}}, \bibinfo
  {author} {\bibfnamefont {D.~A.}\ \bibnamefont {Griffin}}, \ and\ \bibinfo
  {author} {\bibfnamefont {J.}~\bibnamefont {Hafner}},\ }\bibfield  {title}
  {\enquote {\bibinfo {title} {{Analysis of flight MH370 potential debris
  trajectories using ocean observations and numerical model results}},}\
  }\href@noop {} {\bibfield  {journal} {\bibinfo  {journal} {Journal of
  Operational Oceanography}\ }\textbf {\bibinfo {volume} {9}},\ \bibinfo
  {pages} {126--138} (\bibinfo {year} {2016})}\BibitemShut {NoStop}%
\bibitem [{\citenamefont {Nesterov}(2018)}]{Nesterov-18}%
  \BibitemOpen
  \bibfield  {author} {\bibinfo {author} {\bibfnamefont {O.}~\bibnamefont
  {Nesterov}},\ }\bibfield  {title} {\enquote {\bibinfo {title} {{Consideration
  of various aspects in a drift study of MH370 debris}},}\ }\href@noop {}
  {\bibfield  {journal} {\bibinfo  {journal} {Ocean Sci.}\ }\textbf {\bibinfo
  {volume} {14}},\ \bibinfo {pages} {387--402} (\bibinfo {year}
  {2018})}\BibitemShut {NoStop}%
\bibitem [{\citenamefont {Griffin}, \citenamefont {Oke},\ and\ \citenamefont
  {Jones}(2017)}]{Griffin-etal-17}%
  \BibitemOpen
  \bibfield  {author} {\bibinfo {author} {\bibfnamefont {D.}~\bibnamefont
  {Griffin}}, \bibinfo {author} {\bibfnamefont {P.}~\bibnamefont {Oke}}, \ and\
  \bibinfo {author} {\bibfnamefont {E.}~\bibnamefont {Jones}},\ }\href@noop {}
  {\enquote {\bibinfo {title} {{The search for MH370 and ocean surface drift -
  Part II}, 13 april 2017},}\ }\bibinfo {howpublished} {CSIRO Oceans and
  Atmosphere, Australia, Report no. EP172633, prepared for the Australian
  Transport Safety Bureau, available at: https://www.atsb. gov.au} (\bibinfo
  {year} {2017})\BibitemShut {NoStop}%
\bibitem [{\citenamefont {Maximenko}\ \emph {et~al.}(2015)\citenamefont
  {Maximenko}, \citenamefont {Hafner}, \citenamefont {Speidel},\ and\
  \citenamefont {Wang}}]{Maximenko-etal-15}%
  \BibitemOpen
  \bibfield  {author} {\bibinfo {author} {\bibfnamefont {N.}~\bibnamefont
  {Maximenko}}, \bibinfo {author} {\bibfnamefont {J.}~\bibnamefont {Hafner}},
  \bibinfo {author} {\bibfnamefont {J.}~\bibnamefont {Speidel}}, \ and\
  \bibinfo {author} {\bibfnamefont {K.~L.}\ \bibnamefont {Wang}},\ }\href@noop
  {} {\enquote {\bibinfo {title} {{IPRC Ocean Drift Model Simulates MH370 Crash
  Site and Flow Paths}},}\ }\bibinfo {howpublished} {International Pacific
  Research Center, School of Ocean and Earth Science and Technology at the
  University of Hawaii, 4 August 2015, available at:
  http://iprc.soest.hawaii.edu/news/ MH370\_debris/IPRC\_MH370\_News.php}
  (\bibinfo {year} {2015})\BibitemShut {NoStop}%
\bibitem [{\citenamefont {{van Ormondt}}\ and\ \citenamefont
  {Baart}(2015)}]{vanOrmondt-Baart-15}%
  \BibitemOpen
  \bibfield  {author} {\bibinfo {author} {\bibfnamefont {M.}~\bibnamefont {{van
  Ormondt}}}\ and\ \bibinfo {author} {\bibfnamefont {F.}~\bibnamefont
  {Baart}},\ }\href@noop {} {\enquote {\bibinfo {title} {{Aircraft debris MH370
  makes Northern part of the search area more likely}},}\ }\bibinfo
  {howpublished} {Deltares News, 31 July 2015, available at:
  https://www.deltares.nl/en/news/aircraft-
  debris-mh370-makes-northern-part-of-the-search} (\bibinfo {year}
  {2015})\BibitemShut {NoStop}%
\bibitem [{\citenamefont {Jansen}, \citenamefont {Coppin},\ and\ \citenamefont
  {Pinardi}(2016)}]{Jansen-etal-16}%
  \BibitemOpen
  \bibfield  {author} {\bibinfo {author} {\bibfnamefont {E.}~\bibnamefont
  {Jansen}}, \bibinfo {author} {\bibfnamefont {G.}~\bibnamefont {Coppin}}, \
  and\ \bibinfo {author} {\bibfnamefont {N.}~\bibnamefont {Pinardi}},\
  }\bibfield  {title} {\enquote {\bibinfo {title} {{Drift simulation of MH370
  debris using superensemble techniques}},}\ }\href@noop {} {\bibfield
  {journal} {\bibinfo  {journal} {Nat. Hazards Earth Syst. Sci.}\ }\textbf
  {\bibinfo {volume} {16}},\ \bibinfo {pages} {1623--1628} (\bibinfo {year}
  {2016})}\BibitemShut {NoStop}%
\bibitem [{\citenamefont {Durgadoo}\ and\ \citenamefont
  {Biastoch}(2015)}]{Durgadoo-Biastoch-15}%
  \BibitemOpen
  \bibfield  {author} {\bibinfo {author} {\bibfnamefont {J.}~\bibnamefont
  {Durgadoo}}\ and\ \bibinfo {author} {\bibfnamefont {A.}~\bibnamefont
  {Biastoch}},\ }\href@noop {} {\enquote {\bibinfo {title} {{Where is
  MH370?}}}\ }\bibinfo {howpublished} {GEOMAR Helmholtz Centre for Ocean
  Research Kiel, 28 August 2015, available at: http://www.geomar.de} (\bibinfo
  {year} {2015})\BibitemShut {NoStop}%
\bibitem [{\citenamefont {Davey}\ \emph {et~al.}(2016)\citenamefont {Davey},
  \citenamefont {Gordon}, \citenamefont {Holland}, \citenamefont {Rutten},\
  and\ \citenamefont {Williams}}]{Davey-etal-16}%
  \BibitemOpen
  \bibfield  {author} {\bibinfo {author} {\bibfnamefont {S.}~\bibnamefont
  {Davey}}, \bibinfo {author} {\bibfnamefont {N.}~\bibnamefont {Gordon}},
  \bibinfo {author} {\bibfnamefont {I.}~\bibnamefont {Holland}}, \bibinfo
  {author} {\bibfnamefont {M.}~\bibnamefont {Rutten}}, \ and\ \bibinfo {author}
  {\bibfnamefont {J.}~\bibnamefont {Williams}},\ }\href@noop {} {\emph
  {\bibinfo {title} {Bayesian methods in the search for MH370}}},\
  SpringerBriefs in Electrical and Computer Engineering\ (\bibinfo  {publisher}
  {Springer Open},\ \bibinfo {year} {2016})\ p.\ \bibinfo {pages}
  {114}\BibitemShut {NoStop}%
\bibitem [{\citenamefont {{de Deckker}}(2017)}]{deDeckker-17}%
  \BibitemOpen
  \bibfield  {author} {\bibinfo {author} {\bibfnamefont {P.}~\bibnamefont {{de
  Deckker}}},\ }\href@noop {} {\enquote {\bibinfo {title} {{Chemical
  investigations on barnacles found attached to debris from the MH370 aircraft
  found in the Indian Ocean}},}\ }\bibinfo {howpublished} {Appendix F in ATSB
  report: ``The Operational Search for MH370,'' AE-2014-054, 3 October 2017,
  available at: http: //www.atsb.gov.au, last access: 3 October 2017.}
  (\bibinfo {year} {2017})\BibitemShut {NoStop}%
\bibitem [{\citenamefont {Lasota}\ and\ \citenamefont
  {Mackey}(1994)}]{Lasota-Mackey-94}%
  \BibitemOpen
  \bibfield  {author} {\bibinfo {author} {\bibfnamefont {A.}~\bibnamefont
  {Lasota}}\ and\ \bibinfo {author} {\bibfnamefont {M.~C.}\ \bibnamefont
  {Mackey}},\ }\href@noop {} {\emph {\bibinfo {title} {Chaos, Fractals, and
  Noise: Stochastic Aspects of Dynamics}}},\ \bibinfo {edition} {2nd}\ ed.,\
  \bibinfo {series} {Applied Mathematical Sciences}, Vol.~\bibinfo {volume}
  {97}\ (\bibinfo  {publisher} {Springer},\ \bibinfo {address} {New York},\
  \bibinfo {year} {1994})\BibitemShut {NoStop}%
\bibitem [{\citenamefont {Br\'emaud}(1999)}]{Bremaud-99}%
  \BibitemOpen
  \bibfield  {author} {\bibinfo {author} {\bibfnamefont {P.}~\bibnamefont
  {Br\'emaud}},\ }\href@noop {} {\emph {\bibinfo {title} {Markov chains}}},\
  \bibinfo {series} {Gibbs Fields Monte Carlo Simulation Queues, Texts in
  Applied Mathematics}, Vol.~\bibinfo {volume} {31}\ (\bibinfo  {publisher}
  {Springer},\ \bibinfo {address} {New York},\ \bibinfo {year}
  {1999})\BibitemShut {NoStop}%
\bibitem [{\citenamefont {Norris}(1998)}]{Norris-98}%
  \BibitemOpen
  \bibfield  {author} {\bibinfo {author} {\bibfnamefont {J.}~\bibnamefont
  {Norris}},\ }\href@noop {} {\emph {\bibinfo {title} {Markov Chains}}}\
  (\bibinfo  {publisher} {Cambridge University Press},\ \bibinfo {year}
  {1998})\BibitemShut {NoStop}%
\bibitem [{\citenamefont {Beron-Vera}\ \emph {et~al.}(2015)\citenamefont
  {Beron-Vera}, \citenamefont {Olascoaga}, \citenamefont {Haller},
  \citenamefont {Farazmand}, \citenamefont {{Tri\~nanes}},\ and\ \citenamefont
  {Wang}}]{Beron-etal-15}%
  \BibitemOpen
  \bibfield  {author} {\bibinfo {author} {\bibfnamefont {F.~J.}\ \bibnamefont
  {Beron-Vera}}, \bibinfo {author} {\bibfnamefont {M.~J.}\ \bibnamefont
  {Olascoaga}}, \bibinfo {author} {\bibfnamefont {G.}~\bibnamefont {Haller}},
  \bibinfo {author} {\bibfnamefont {M.}~\bibnamefont {Farazmand}}, \bibinfo
  {author} {\bibfnamefont {J.}~\bibnamefont {{Tri\~nanes}}}, \ and\ \bibinfo
  {author} {\bibfnamefont {Y.}~\bibnamefont {Wang}},\ }\bibfield  {title}
  {\enquote {\bibinfo {title} {{Dissipative inertial transport patterns near
  coherent Lagrangian eddies in the ocean}},}\ }\href {\doibase
  10.1063/1.4928693} {\bibfield  {journal} {\bibinfo  {journal} {Chaos}\
  }\textbf {\bibinfo {volume} {25}},\ \bibinfo {pages} {087412} (\bibinfo
  {year} {2015})}\BibitemShut {NoStop}%
\bibitem [{\citenamefont {Beron-Vera}, \citenamefont {Olascoaga},\ and\
  \citenamefont {Lumpkin}(2016)}]{Beron-etal-16}%
  \BibitemOpen
  \bibfield  {author} {\bibinfo {author} {\bibfnamefont {F.~J.}\ \bibnamefont
  {Beron-Vera}}, \bibinfo {author} {\bibfnamefont {M.~J.}\ \bibnamefont
  {Olascoaga}}, \ and\ \bibinfo {author} {\bibfnamefont {R.}~\bibnamefont
  {Lumpkin}},\ }\bibfield  {title} {\enquote {\bibinfo {title} {Inertia-induced
  accumulation of flotsam in the subtropical gyres},}\ }\href@noop {}
  {\bibfield  {journal} {\bibinfo  {journal} {Geophys. Res. Lett.}\ }\textbf
  {\bibinfo {volume} {43}},\ \bibinfo {pages} {12228--12233} (\bibinfo {year}
  {2016})}\BibitemShut {NoStop}%
\bibitem [{\citenamefont {Ulam}(1960)}]{Ulam-60}%
  \BibitemOpen
  \bibfield  {author} {\bibinfo {author} {\bibfnamefont {S.~M.}\ \bibnamefont
  {Ulam}},\ }\href@noop {} {\emph {\bibinfo {title} {A Collection of
  Mathematical Problems}}},\ Interscience tracts in pure and applied
  mathematics\ (\bibinfo  {publisher} {Interscience},\ \bibinfo {year}
  {1960})\BibitemShut {NoStop}%
\bibitem [{\citenamefont {Kov\'acs}\ and\ \citenamefont
  {T\'el}(1989)}]{Kovacs-Tel-89}%
  \BibitemOpen
  \bibfield  {author} {\bibinfo {author} {\bibfnamefont {Z.}~\bibnamefont
  {Kov\'acs}}\ and\ \bibinfo {author} {\bibfnamefont {T.}~\bibnamefont
  {T\'el}},\ }\bibfield  {title} {\enquote {\bibinfo {title} {{Scaling in
  multifractals: Discretization of an eigenvalue problem}},}\ }\href@noop {}
  {\bibfield  {journal} {\bibinfo  {journal} {Phys. Rev. A}\ }\textbf {\bibinfo
  {volume} {40}},\ \bibinfo {pages} {4641--4646} (\bibinfo {year}
  {1989})}\BibitemShut {NoStop}%
\bibitem [{\citenamefont {Koltai}(2010)}]{Koltai-10}%
  \BibitemOpen
  \bibfield  {author} {\bibinfo {author} {\bibfnamefont {P.}~\bibnamefont
  {Koltai}},\ }\emph {\bibinfo {title} {Efficient approximation methods for the
  global long-term behavior of dynamical systems -- Theory, algorithms and
  examples}},\ \href@noop {} {Ph.D. thesis},\ \bibinfo  {school} {Technical
  University of Munich} (\bibinfo {year} {2010})\BibitemShut {NoStop}%
\bibitem [{\citenamefont {Miron}\ \emph {et~al.}(2018)\citenamefont {Miron},
  \citenamefont {Beron-Vera}, \citenamefont {Olascoaga}, \citenamefont
  {Froyland}, \citenamefont {P\'erez-Brunius},\ and\ \citenamefont
  {Sheinbaum}}]{Miron-etal-18a}%
  \BibitemOpen
  \bibfield  {author} {\bibinfo {author} {\bibfnamefont {P.}~\bibnamefont
  {Miron}}, \bibinfo {author} {\bibfnamefont {F.~J.}\ \bibnamefont
  {Beron-Vera}}, \bibinfo {author} {\bibfnamefont {M.~J.}\ \bibnamefont
  {Olascoaga}}, \bibinfo {author} {\bibfnamefont {G.}~\bibnamefont {Froyland}},
  \bibinfo {author} {\bibfnamefont {P.}~\bibnamefont {P\'erez-Brunius}}, \ and\
  \bibinfo {author} {\bibfnamefont {J.}~\bibnamefont {Sheinbaum}},\ }\bibfield
  {title} {\enquote {\bibinfo {title} {{Lagrangian geography of the deep Gulf
  of Mexico}},}\ }\href@noop {} {\bibfield  {journal} {\bibinfo  {journal} {J.
  Phys. Oceanogr.}\ }\textbf {\bibinfo {volume} {doi:0.1175/JPO-D-18-0073.1}}
  (\bibinfo {year} {2018})}\BibitemShut {NoStop}%
\bibitem [{\citenamefont {Schott}\ and\ \citenamefont {{McCreary,
  Jr.}}(2001)}]{Schott-McCreary-01}%
  \BibitemOpen
  \bibfield  {author} {\bibinfo {author} {\bibfnamefont {F.~A.}\ \bibnamefont
  {Schott}}\ and\ \bibinfo {author} {\bibfnamefont {J.~P.}\ \bibnamefont
  {{McCreary, Jr.}}},\ }\bibfield  {title} {\enquote {\bibinfo {title} {{The
  monsoon circulation of the Indian Ocean}},}\ }\href@noop {} {\bibfield
  {journal} {\bibinfo  {journal} {Progress In Oceanography}\ }\textbf {\bibinfo
  {volume} {51}},\ \bibinfo {pages} {1--123} (\bibinfo {year}
  {2001})}\BibitemShut {NoStop}%
\bibitem [{\citenamefont {Lumpkin}\ and\ \citenamefont
  {Pazos}(2007)}]{Lumpkin-Pazos-07}%
  \BibitemOpen
  \bibfield  {author} {\bibinfo {author} {\bibfnamefont {R.}~\bibnamefont
  {Lumpkin}}\ and\ \bibinfo {author} {\bibfnamefont {M.}~\bibnamefont
  {Pazos}},\ }\bibfield  {title} {\enquote {\bibinfo {title} {{Measuring
  surface currents with Surface Velocity Program drifters: the instrument, its
  data and some recent results}},}\ }in\ \href@noop {} {\emph {\bibinfo
  {booktitle} {Lagrangian Analysis and Prediction of Coastal and Ocean
  Dynamics}}},\ \bibinfo {editor} {edited by\ \bibinfo {editor} {\bibfnamefont
  {A.}~\bibnamefont {Griffa}}, \bibinfo {editor} {\bibfnamefont {A.~D.}\
  \bibnamefont {Kirwan}}, \bibinfo {editor} {\bibfnamefont {A.}~\bibnamefont
  {Mariano}}, \bibinfo {editor} {\bibfnamefont {T.}~\bibnamefont
  {\"{O}zg\"{o}kmen}}, \ and\ \bibinfo {editor} {\bibfnamefont
  {T.}~\bibnamefont {Rossby}}}\ (\bibinfo  {publisher} {Cambridge University
  Press},\ \bibinfo {year} {2007})\ Chap.~\bibinfo {chapter} {2}, pp.\ \bibinfo
  {pages} {39--67}\BibitemShut {NoStop}%
\bibitem [{\citenamefont {Niiler}\ and\ \citenamefont
  {Paduan}(1995)}]{Niiler-Paduan-95}%
  \BibitemOpen
  \bibfield  {author} {\bibinfo {author} {\bibfnamefont {P.~P.}\ \bibnamefont
  {Niiler}}\ and\ \bibinfo {author} {\bibfnamefont {J.~D.}\ \bibnamefont
  {Paduan}},\ }\bibfield  {title} {\enquote {\bibinfo {title} {{Wind-driven
  Motions in the northeastern Pacific as measured by Lagrangian drifters}},}\
  }\href@noop {} {\bibfield  {journal} {\bibinfo  {journal} {J. Phys.
  Oceanogr.}\ }\textbf {\bibinfo {volume} {25}},\ \bibinfo {pages} {2819--2830}
  (\bibinfo {year} {1995})}\BibitemShut {NoStop}%
\bibitem [{\citenamefont {{Lumpkin}}\ \emph {et~al.}(2012)\citenamefont
  {{Lumpkin}}, \citenamefont {{Grodsky}}, \citenamefont {{Centurioni}},
  \citenamefont {{Rio}}, \citenamefont {{Carton}},\ and\ \citenamefont
  {{Lee}}}]{Lumpkin-etal-12}%
  \BibitemOpen
  \bibfield  {author} {\bibinfo {author} {\bibfnamefont {R.}~\bibnamefont
  {{Lumpkin}}}, \bibinfo {author} {\bibfnamefont {S.~A.}\ \bibnamefont
  {{Grodsky}}}, \bibinfo {author} {\bibfnamefont {L.}~\bibnamefont
  {{Centurioni}}}, \bibinfo {author} {\bibfnamefont {M.-H.}\ \bibnamefont
  {{Rio}}}, \bibinfo {author} {\bibfnamefont {J.~A.}\ \bibnamefont {{Carton}}},
  \ and\ \bibinfo {author} {\bibfnamefont {D.}~\bibnamefont {{Lee}}},\
  }\bibfield  {title} {\enquote {\bibinfo {title} {Removing spurious
  low-frequency variability in drifter velocities},}\ }\href@noop {} {\bibfield
   {journal} {\bibinfo  {journal} {J. Atm. Oce. Tech.}\ }\textbf {\bibinfo
  {volume} {30}},\ \bibinfo {pages} {353--360} (\bibinfo {year}
  {2012})}\BibitemShut {NoStop}%
\bibitem [{\citenamefont {Miron}\ \emph {et~al.}(2017)\citenamefont {Miron},
  \citenamefont {Beron-Vera}, \citenamefont {Olascoaga}, \citenamefont
  {Sheinbaum}, \citenamefont {P\'erez-Brunius},\ and\ \citenamefont
  {Froyland}}]{Miron-etal-17}%
  \BibitemOpen
  \bibfield  {author} {\bibinfo {author} {\bibfnamefont {P.}~\bibnamefont
  {Miron}}, \bibinfo {author} {\bibfnamefont {F.~J.}\ \bibnamefont
  {Beron-Vera}}, \bibinfo {author} {\bibfnamefont {M.~J.}\ \bibnamefont
  {Olascoaga}}, \bibinfo {author} {\bibfnamefont {J.}~\bibnamefont
  {Sheinbaum}}, \bibinfo {author} {\bibfnamefont {P.}~\bibnamefont
  {P\'erez-Brunius}}, \ and\ \bibinfo {author} {\bibfnamefont {G.}~\bibnamefont
  {Froyland}},\ }\bibfield  {title} {\enquote {\bibinfo {title} {{Lagrangian
  dynamical geography of the Gulf of Mexico}},}\ }\href {\doibase
  10.1038/s41598-017-07177-w} {\bibfield  {journal} {\bibinfo  {journal}
  {Scientific Reports}\ }\textbf {\bibinfo {volume} {7}},\ \bibinfo {pages}
  {7021} (\bibinfo {year} {2017})}\BibitemShut {NoStop}%
\bibitem [{\citenamefont {Olascoaga}\ \emph {et~al.}(2018)\citenamefont
  {Olascoaga}, \citenamefont {Miron}, \citenamefont {Paris}, \citenamefont
  {P\'erez-Brunius}, \citenamefont {P\'erez-Portela}, \citenamefont {Smith},\
  and\ \citenamefont {Vaz}}]{Olascoaga-etal-18}%
  \BibitemOpen
  \bibfield  {author} {\bibinfo {author} {\bibfnamefont {M.~J.}\ \bibnamefont
  {Olascoaga}}, \bibinfo {author} {\bibfnamefont {P.}~\bibnamefont {Miron}},
  \bibinfo {author} {\bibfnamefont {C.}~\bibnamefont {Paris}}, \bibinfo
  {author} {\bibfnamefont {P.}~\bibnamefont {P\'erez-Brunius}}, \bibinfo
  {author} {\bibfnamefont {R.}~\bibnamefont {P\'erez-Portela}}, \bibinfo
  {author} {\bibfnamefont {R.~H.}\ \bibnamefont {Smith}}, \ and\ \bibinfo
  {author} {\bibfnamefont {A.}~\bibnamefont {Vaz}},\ }\bibfield  {title}
  {\enquote {\bibinfo {title} {{Connectivity of Pulley Ridge with remote
  locations as inferred from satellite-tracked drifter trajectories}},}\
  }\href@noop {} {\bibfield  {journal} {\bibinfo  {journal} {Journal of
  Geophysical Research}\ }\textbf {\bibinfo {volume} {123}},\ \bibinfo {pages}
  {5742--5750} (\bibinfo {year} {2018})}\BibitemShut {NoStop}%
\bibitem [{\citenamefont {Froyland}\ and\ \citenamefont
  {Padberg}(2009)}]{Froyland-Padberg-09}%
  \BibitemOpen
  \bibfield  {author} {\bibinfo {author} {\bibfnamefont {G.}~\bibnamefont
  {Froyland}}\ and\ \bibinfo {author} {\bibfnamefont {K.}~\bibnamefont
  {Padberg}},\ }\bibfield  {title} {\enquote {\bibinfo {title}
  {Almost-invariant sets and invariant manifolds --- connecting probabilistic
  and geometric descriptions of coherent structures in flows},}\ }\href@noop {}
  {\bibfield  {journal} {\bibinfo  {journal} {Physica D}\ }\textbf {\bibinfo
  {volume} {238}},\ \bibinfo {pages} {1507--1523} (\bibinfo {year}
  {2009})}\BibitemShut {NoStop}%
\bibitem [{\citenamefont {LaCasce}(2008)}]{Lacasce-08}%
  \BibitemOpen
  \bibfield  {author} {\bibinfo {author} {\bibfnamefont {J.~H.}\ \bibnamefont
  {LaCasce}},\ }\bibfield  {title} {\enquote {\bibinfo {title} {{Statistics
  from Lagrangian observations}},}\ }\href@noop {} {\bibfield  {journal}
  {\bibinfo  {journal} {Progr. Oceanogr.}\ }\textbf {\bibinfo {volume} {77}},\
  \bibinfo {pages} {1--29} (\bibinfo {year} {2008})}\BibitemShut {NoStop}%
\bibitem [{\citenamefont {Maximenko}, \citenamefont {Hafner},\ and\
  \citenamefont {Niiler}(2012)}]{Maximenko-etal-12}%
  \BibitemOpen
  \bibfield  {author} {\bibinfo {author} {\bibfnamefont {A.~N.}\ \bibnamefont
  {Maximenko}}, \bibinfo {author} {\bibfnamefont {J.}~\bibnamefont {Hafner}}, \
  and\ \bibinfo {author} {\bibfnamefont {P.}~\bibnamefont {Niiler}},\
  }\bibfield  {title} {\enquote {\bibinfo {title} {{Pathways of marine debris
  derived from trajectories of Lagrangian drifters}},}\ }\href@noop {}
  {\bibfield  {journal} {\bibinfo  {journal} {Mar. Pollut. Bull.}\ }\textbf
  {\bibinfo {volume} {65}},\ \bibinfo {pages} {51--62} (\bibinfo {year}
  {2012})}\BibitemShut {NoStop}%
\bibitem [{\citenamefont {McAdam}\ and\ \citenamefont {van
  Sebille}(2018)}]{McAdam-vanSebille-18}%
  \BibitemOpen
  \bibfield  {author} {\bibinfo {author} {\bibfnamefont {R.}~\bibnamefont
  {McAdam}}\ and\ \bibinfo {author} {\bibfnamefont {E.}~\bibnamefont {van
  Sebille}},\ }\bibfield  {title} {\enquote {\bibinfo {title} {Surface
  connectivity and interocean exchanges from drifter-based transition
  matrices},}\ }\href@noop {} {\bibfield  {journal} {\bibinfo  {journal}
  {Journal of Geophysical Research: Oceans}\ }\textbf {\bibinfo {volume}
  {123}},\ \bibinfo {pages} {514--532} (\bibinfo {year} {2018})}\BibitemShut
  {NoStop}%
\bibitem [{\citenamefont {{van Sebille}}, \citenamefont {England},\ and\
  \citenamefont {Froyland}(2012)}]{vanSebille-etal-12}%
  \BibitemOpen
  \bibfield  {author} {\bibinfo {author} {\bibfnamefont {E.}~\bibnamefont {{van
  Sebille}}}, \bibinfo {author} {\bibfnamefont {E.~H.}\ \bibnamefont
  {England}}, \ and\ \bibinfo {author} {\bibfnamefont {G.}~\bibnamefont
  {Froyland}},\ }\bibfield  {title} {\enquote {\bibinfo {title} {Origin,
  dynamics and evolution of ocean garbage patches from observed surface
  drifters},}\ }\href@noop {} {\bibfield  {journal} {\bibinfo  {journal}
  {Environ. Res. Lett.}\ }\textbf {\bibinfo {volume} {7}},\ \bibinfo {pages}
  {044040} (\bibinfo {year} {2012})}\BibitemShut {NoStop}%
\bibitem [{\citenamefont {Froyland}, \citenamefont {Stuart},\ and\
  \citenamefont {{van Sebille}}(2014)}]{Froyland-etal-14}%
  \BibitemOpen
  \bibfield  {author} {\bibinfo {author} {\bibfnamefont {G.}~\bibnamefont
  {Froyland}}, \bibinfo {author} {\bibfnamefont {R.~M.}\ \bibnamefont
  {Stuart}}, \ and\ \bibinfo {author} {\bibfnamefont {E.}~\bibnamefont {{van
  Sebille}}},\ }\bibfield  {title} {\enquote {\bibinfo {title} {How
  well-connected is the surface of the global ocean?}}\ }\href@noop {}
  {\bibfield  {journal} {\bibinfo  {journal} {Chaos}\ }\textbf {\bibinfo
  {volume} {24}},\ \bibinfo {pages} {033126} (\bibinfo {year}
  {2014})}\BibitemShut {NoStop}%
\bibitem [{\citenamefont {Hsu}(1987)}]{Hsu-87}%
  \BibitemOpen
  \bibfield  {author} {\bibinfo {author} {\bibfnamefont {C.~S.}\ \bibnamefont
  {Hsu}},\ }\href@noop {} {\emph {\bibinfo {title} {Cell-to-cell mapping. A
  Method of Global Analysis for Nonlinear Systems}}},\ \bibinfo {series}
  {Applied Mathematical Sciences}, Vol.~\bibinfo {volume} {64}\ (\bibinfo
  {publisher} {Springer-Verlag},\ \bibinfo {address} {New York},\ \bibinfo
  {year} {1987})\ p.\ \bibinfo {pages} {354}\BibitemShut {NoStop}%
\bibitem [{\citenamefont {Dellnitz}\ and\ \citenamefont
  {Junge}(1999)}]{Dellnitz-Junge-99}%
  \BibitemOpen
  \bibfield  {author} {\bibinfo {author} {\bibfnamefont {M.}~\bibnamefont
  {Dellnitz}}\ and\ \bibinfo {author} {\bibfnamefont {O.}~\bibnamefont
  {Junge}},\ }\bibfield  {title} {\enquote {\bibinfo {title} {On the
  approximation of complicated dynamical behavior},}\ }\href@noop {} {\bibfield
   {journal} {\bibinfo  {journal} {SIAM J. Numer. Anal.}\ }\textbf {\bibinfo
  {volume} {36}},\ \bibinfo {pages} {491--515} (\bibinfo {year}
  {1999})}\BibitemShut {NoStop}%
\bibitem [{\citenamefont {Froyland}(2005)}]{Froyland-05}%
  \BibitemOpen
  \bibfield  {author} {\bibinfo {author} {\bibfnamefont {G.}~\bibnamefont
  {Froyland}},\ }\bibfield  {title} {\enquote {\bibinfo {title} {Statistically
  optimal almost-invariant sets},}\ }\href@noop {} {\bibfield  {journal}
  {\bibinfo  {journal} {Physica D}\ }\textbf {\bibinfo {volume} {200}},\
  \bibinfo {pages} {205--219} (\bibinfo {year} {2005})}\BibitemShut {NoStop}%
\bibitem [{\citenamefont {Horn}\ and\ \citenamefont
  {Johnson}(1990)}]{Horn-Johnson-90}%
  \BibitemOpen
  \bibfield  {author} {\bibinfo {author} {\bibfnamefont {R.~A.}\ \bibnamefont
  {Horn}}\ and\ \bibinfo {author} {\bibfnamefont {C.~R.}\ \bibnamefont
  {Johnson}},\ }\href@noop {} {\emph {\bibinfo {title} {Matrix Analysis}}}\
  (\bibinfo  {publisher} {Cambridge University Press},\ \bibinfo {year}
  {1990})\BibitemShut {NoStop}%
\bibitem [{\citenamefont {Koltai}(2011)}]{Koltai-11}%
  \BibitemOpen
  \bibfield  {author} {\bibinfo {author} {\bibfnamefont {P.}~\bibnamefont
  {Koltai}},\ }\bibfield  {title} {\enquote {\bibinfo {title} {A stochastic
  approach for computing the domain of attraction without trajectory
  simulation},}\ }in\ \href@noop {} {\emph {\bibinfo {booktitle} {Dynamical
  Systems, Differential Equations and Applications, 8th AIMS Conference.
  Suppl.}}},\ Vol.~\bibinfo {volume} {2}\ (\bibinfo {year} {2011})\ pp.\
  \bibinfo {pages} {854--863}\BibitemShut {NoStop}%
\bibitem [{\citenamefont {Beal}\ \emph {et~al.}(2011)\citenamefont {Beal},
  \citenamefont {{de Ruijter}}, \citenamefont {Biastoch}, \citenamefont
  {Zahn},\ and\ \citenamefont {{SCOR/WCRP/IAPSO Working Group
  136}}}]{Beal-etal-11}%
  \BibitemOpen
  \bibfield  {author} {\bibinfo {author} {\bibfnamefont {L.~M.}\ \bibnamefont
  {Beal}}, \bibinfo {author} {\bibfnamefont {W.~P.~M.}\ \bibnamefont {{de
  Ruijter}}}, \bibinfo {author} {\bibfnamefont {A.}~\bibnamefont {Biastoch}},
  \bibinfo {author} {\bibfnamefont {R.}~\bibnamefont {Zahn}}, \ and\ \bibinfo
  {author} {\bibnamefont {{SCOR/WCRP/IAPSO Working Group 136}}},\ }\bibfield
  {title} {\enquote {\bibinfo {title} {{On the role of the Agulhas system in
  ocean circulation and climate}},}\ }\href@noop {} {\bibfield  {journal}
  {\bibinfo  {journal} {Nature}\ }\textbf {\bibinfo {volume} {472}},\ \bibinfo
  {pages} {429--436} (\bibinfo {year} {2011})}\BibitemShut {NoStop}%
\bibitem [{\citenamefont {Bolstad}\ and\ \citenamefont
  {Curran}(2016)}]{Bolstad-Curran-16}%
  \BibitemOpen
  \bibfield  {author} {\bibinfo {author} {\bibfnamefont {W.~M.}\ \bibnamefont
  {Bolstad}}\ and\ \bibinfo {author} {\bibfnamefont {J.~M.}\ \bibnamefont
  {Curran}},\ }\href@noop {} {\emph {\bibinfo {title} {Introduction to Bayesian
  statistics}}}\ (\bibinfo  {publisher} {John Wiley \& Sons},\ \bibinfo {year}
  {2016})\BibitemShut {NoStop}%
\bibitem [{\citenamefont {Dijkstra}(1959)}]{Dijkstra-59}%
  \BibitemOpen
  \bibfield  {author} {\bibinfo {author} {\bibfnamefont {E.~W.}\ \bibnamefont
  {Dijkstra}},\ }\bibfield  {title} {\enquote {\bibinfo {title} {A note on two
  problems in connexion with graphs},}\ }\href@noop {} {\bibfield  {journal}
  {\bibinfo  {journal} {Numerische Mathematik}\ }\textbf {\bibinfo {volume}
  {1}},\ \bibinfo {pages} {269--271} (\bibinfo {year} {1959})}\BibitemShut
  {NoStop}%
\bibitem [{\citenamefont {Floyd}(1962)}]{Floyd-62}%
  \BibitemOpen
  \bibfield  {author} {\bibinfo {author} {\bibfnamefont {R.~W.}\ \bibnamefont
  {Floyd}},\ }\bibfield  {title} {\enquote {\bibinfo {title} {Algorithm 97:
  shortest path},}\ }\href@noop {} {\bibfield  {journal} {\bibinfo  {journal}
  {Communications of the ACM}\ }\textbf {\bibinfo {volume} {5}},\ \bibinfo
  {pages} {345} (\bibinfo {year} {1962})}\BibitemShut {NoStop}%
\bibitem [{\citenamefont {Ser-Giacomi}\ \emph {et~al.}(2015)\citenamefont
  {Ser-Giacomi}, \citenamefont {Vasile}, \citenamefont
  {Hern{\'a}ndez-Garc{\'\i}a},\ and\ \citenamefont
  {L{\'o}pez}}]{SerGiacomi-etal-15b}%
  \BibitemOpen
  \bibfield  {author} {\bibinfo {author} {\bibfnamefont {E.}~\bibnamefont
  {Ser-Giacomi}}, \bibinfo {author} {\bibfnamefont {R.}~\bibnamefont {Vasile}},
  \bibinfo {author} {\bibfnamefont {E.}~\bibnamefont
  {Hern{\'a}ndez-Garc{\'\i}a}}, \ and\ \bibinfo {author} {\bibfnamefont
  {C.}~\bibnamefont {L{\'o}pez}},\ }\bibfield  {title} {\enquote {\bibinfo
  {title} {Most probable paths in temporal weighted networks: An application to
  ocean transport},}\ }\href@noop {} {\bibfield  {journal} {\bibinfo  {journal}
  {Physical review E}\ }\textbf {\bibinfo {volume} {92}},\ \bibinfo {pages}
  {012818} (\bibinfo {year} {2015})}\BibitemShut {NoStop}%
\bibitem [{\citenamefont {{Australian Transport Safety Bureau}}()}]{ATSB-14}%
  \BibitemOpen
  \bibfield  {author} {\bibinfo {author} {\bibnamefont {{Australian Transport
  Safety Bureau}}},\ }\href@noop {} {\enquote {\bibinfo {title} {Mh370 -
  definition of underwater search areas, 26 june 2014},}\ }\bibinfo
  {howpublished} {available at: http://www.atsb. gov.au}\BibitemShut {NoStop}%
\bibitem [{\citenamefont {Cartwright}\ \emph {et~al.}(2010)\citenamefont
  {Cartwright}, \citenamefont {Feudel}, \citenamefont {K\'arolyi},
  \citenamefont {{de Moura}}, \citenamefont {Piro},\ and\ \citenamefont
  {T\'el}}]{Cartwright-etal-10}%
  \BibitemOpen
  \bibfield  {author} {\bibinfo {author} {\bibfnamefont {J.~H.~E.}\
  \bibnamefont {Cartwright}}, \bibinfo {author} {\bibfnamefont
  {U.}~\bibnamefont {Feudel}}, \bibinfo {author} {\bibfnamefont
  {G.}~\bibnamefont {K\'arolyi}}, \bibinfo {author} {\bibfnamefont
  {A.}~\bibnamefont {{de Moura}}}, \bibinfo {author} {\bibfnamefont
  {O.}~\bibnamefont {Piro}}, \ and\ \bibinfo {author} {\bibfnamefont
  {T.}~\bibnamefont {T\'el}},\ }\bibfield  {title} {\enquote {\bibinfo {title}
  {Dynamics of finite-size particles in chaotic fluid flows},}\ }in\ \href@noop
  {} {\emph {\bibinfo {booktitle} {Nonlinear Dynamics and Chaos: Advances and
  Perspectives}}},\ \bibinfo {editor} {edited by\ \bibinfo {editor}
  {\bibnamefont {{M. Thiel et al.}}}}\ (\bibinfo  {publisher} {Springer-Verlag
  Berlin Heidelberg},\ \bibinfo {year} {2010})\ pp.\ \bibinfo {pages}
  {51--87}\BibitemShut {NoStop}%
\bibitem [{\citenamefont {Bagtzoglou}\ and\ \citenamefont
  {Atmadja}(2005)}]{Bagtzoglou-Atmadja-05}%
  \BibitemOpen
  \bibfield  {author} {\bibinfo {author} {\bibfnamefont {A.~C.}\ \bibnamefont
  {Bagtzoglou}}\ and\ \bibinfo {author} {\bibfnamefont {J.}~\bibnamefont
  {Atmadja}},\ }\enquote {\bibinfo {title} {Mathematical methods for hydrologic
  inversion: The case of pollution source identification},}\ in\ \href@noop {}
  {\emph {\bibinfo {booktitle} {Water Pollution: Environmental Impact
  Assessment of Recycled Wastes on Surface and Ground Waters; Engineering
  Modeling and Sustainability}}},\ \bibinfo {editor} {edited by\ \bibinfo
  {editor} {\bibfnamefont {T.~A.}\ \bibnamefont {Kassim}}}\ (\bibinfo
  {publisher} {Springer Berlin Heidelberg},\ \bibinfo {address} {Berlin,
  Heidelberg},\ \bibinfo {year} {2005})\ pp.\ \bibinfo {pages}
  {65--96}\BibitemShut {NoStop}%
\bibitem [{\citenamefont {Olascoaga}\ \emph {et~al.}(2008)\citenamefont
  {Olascoaga}, \citenamefont {Beron-Vera}, \citenamefont {Brand},\ and\
  \citenamefont {Ko\c{c}ak}}]{Olascoaga-etal-08}%
  \BibitemOpen
  \bibfield  {author} {\bibinfo {author} {\bibfnamefont {M.~J.}\ \bibnamefont
  {Olascoaga}}, \bibinfo {author} {\bibfnamefont {F.~J.}\ \bibnamefont
  {Beron-Vera}}, \bibinfo {author} {\bibfnamefont {L.~E.}\ \bibnamefont
  {Brand}}, \ and\ \bibinfo {author} {\bibfnamefont {H.}~\bibnamefont
  {Ko\c{c}ak}},\ }\bibfield  {title} {\enquote {\bibinfo {title} {{Tracing the
  early development of harmful algal blooms on the West Florida Shelf with the
  aid of Lagrangian coherent structures}},}\ }\href {\doibase
  10.1029/2007JC004533} {\bibfield  {journal} {\bibinfo  {journal} {J. Geophys.
  Res.}\ }\textbf {\bibinfo {volume} {113}},\ \bibinfo {pages} {C12014}
  (\bibinfo {year} {2008})}\BibitemShut {NoStop}%
\bibitem [{\citenamefont {{Gautama}}\ \emph {et~al.}(2016)\citenamefont
  {{Gautama}}, \citenamefont {{Mercier}}, \citenamefont {{Fablet}},\ and\
  \citenamefont {{Longepe}}}]{Gautama-etal-16}%
  \BibitemOpen
  \bibfield  {author} {\bibinfo {author} {\bibfnamefont {B.~G.}\ \bibnamefont
  {{Gautama}}}, \bibinfo {author} {\bibfnamefont {G.}~\bibnamefont
  {{Mercier}}}, \bibinfo {author} {\bibfnamefont {R.}~\bibnamefont {{Fablet}}},
  \ and\ \bibinfo {author} {\bibfnamefont {N.}~\bibnamefont {{Longepe}}},\
  }\bibfield  {title} {\enquote {\bibinfo {title} {{Lagrangian-based
  Backtracking of Oil Spill Dynamics from SAR Images: Application to Montara
  Case}},}\ }in\ \href@noop {} {\emph {\bibinfo {booktitle} {Living Planet
  Symposium}}},\ \bibinfo {series} {ESA Special Publication}, Vol.\ \bibinfo
  {volume} {740}\ (\bibinfo {year} {2016})\ p.\ \bibinfo {pages}
  {214}\BibitemShut {NoStop}%
\bibitem [{\citenamefont {Hourdin}\ and\ \citenamefont
  {Talagrand}(2006)}]{Hourdin-Talagrand-06}%
  \BibitemOpen
  \bibfield  {author} {\bibinfo {author} {\bibfnamefont {F.}~\bibnamefont
  {Hourdin}}\ and\ \bibinfo {author} {\bibfnamefont {O.}~\bibnamefont
  {Talagrand}},\ }\bibfield  {title} {\enquote {\bibinfo {title} {{Eulerian
  backtracking of atmospheric tracers. I: Adjoint derivation and
  parametrization of subgrid-scale transport}},}\ }\href@noop {} {\bibfield
  {journal} {\bibinfo  {journal} {Q. J. R. Meteorol. Soc.}\ }\textbf {\bibinfo
  {volume} {132}},\ \bibinfo {pages} {567--583} (\bibinfo {year}
  {2006})}\BibitemShut {NoStop}%
\bibitem [{\citenamefont {Rao}(2007)}]{Rao-07}%
  \BibitemOpen
  \bibfield  {author} {\bibinfo {author} {\bibfnamefont {K.~S.}\ \bibnamefont
  {Rao}},\ }\bibfield  {title} {\enquote {\bibinfo {title} {Source estimation
  methods for atmospheric dispersion},}\ }\href {\doibase
  https://doi.org/10.1016/j.atmosenv.2007.04.064} {\bibfield  {journal}
  {\bibinfo  {journal} {Atmospheric Environment}\ }\textbf {\bibinfo {volume}
  {41}},\ \bibinfo {pages} {6964 -- 6973} (\bibinfo {year} {2007})}\BibitemShut
  {NoStop}%
\end{thebibliography}

\end{document}